\documentclass[
reprint,
amsmath,
amssymb,
floatfix,
superscriptaddress,
onecolumn,
aps,
prl
]{revtex4-2}

\usepackage{dcolumn}
\usepackage{ulem}
\usepackage{bm}
\usepackage{color}
\usepackage{amssymb}
\usepackage{amsmath}
\usepackage{xfrac}
\usepackage{mathrsfs}
\usepackage{comment}
\usepackage{graphicx}

\renewcommand{\thesection}{\arabic{section}}
\renewcommand{\thesubsection}{\thesection.\arabic{subsection}}
\renewcommand{\thesubsubsection}{\thesubsection.\arabic{subsubsection}}

\def\br{{\bm{r}}}
\def\bv{{\bm{v}}}
\def\tbr{{\tilde{\bm{r}}}}
\def\tbv{{\tilde{\bm{v}}}}
\def\Pe{{\rm{Pe}}}
\def\bu{{\bm{\hat{u}}}}

\def\ttt{{\tilde{t}}}

\def\tbk{{\tilde{\bm{k}}}}

\begin{document}

\title[Inertial chiral active Brownian particle]{Inertia--chirality interplay in active Brownian motion: exact dynamics and phase maps}

\author{Anweshika Pattanayak}
\email{anweshika.pattanayak@tifr.res.in}
\affiliation{Department of Physical Sciences, Indian Institute of Science Education and Research Mohali, Sector 81, Knowledge City, S. A. S. Nagar, Manauli PO 140306, India}
\affiliation{Department of Theoretical Physics, Tata Institute of Fundamental Research, \\Homi Bhabha Road, Mumbai 400005, India}

\author{Sandip Roy}
\email{mp16001@iisermohali.ac.in}
\affiliation{Department of Physical Sciences, Indian Institute of Science Education and Research Mohali, Sector 81, Knowledge City, S. A. S. Nagar, Manauli PO 140306, India}

\author{Abhishek Chaudhuri}
\email{abhishek@iisermohali.ac.in}
\affiliation{Department of Physical Sciences, Indian Institute of Science Education and Research Mohali, Sector 81, Knowledge City, S. A. S. Nagar, Manauli PO 140306, India}

\begin{abstract}
We present an exact, time-resolved theory for a two-dimensional chiral active Brownian particle (cABP) with translational inertia. Using a Laplace-transform moment hierarchy, we derive closed-form expressions for the mean velocity, velocity--orientation projections, velocity autocorrelation, mean-squared velocity, mean-squared displacement, and the fourth moment of velocity. These results agree quantitatively with simulations over all masses, activities, and chiralities. We show that the velocity autocorrelation factorizes into an inertial envelope and a chiral envelope.
Despite rich transients in the velocity sector, the long-time positional diffusion equals the overdamped cABP value, 
independent of mass. From the steady mean-squared velocity, we define a kinetic temperature and a modified fluctuation--dissipation relation whose violation vanishes in two limits: large mass or large chirality, identifying chirality as an additional route to equilibrium-like behavior. The steady-state velocity excess kurtosis gives a phase map
that exhibits a (Gaussian-like)--active(bimodal)--(Gaussian-like) re-entrance with mass; chirality \emph{confines} activity and shrinks the active sector. A narrow positive-kurtosis window emerges at large mass and intermediate chirality, with analytic boundaries consistent with the heavy-mass asymptote. 
\end{abstract}

\maketitle

\section{Introduction}
The study of active Brownian systems has gained significant attention in recent years. These systems continuously convert internal or external energy into mechanical work, generating self-propulsion forces that drive persistent motion ~\cite{romanczuk2012active,Marchetti2013hydrodynamics,bechinger2016active,stenhammar2014phase,das2020introduction,bernheim2018living,bar2020self,deseigne2010collective,kumar2014flocking,takatori2014swim}. As a result, they inherently break the condition of local detailed balance, placing them far from equilibrium. 
Active matter spans molecular to macroscopic scales: from motor proteins inside cells~\cite{needleman2017active,shelley2016dynamics}, collective cell and tissue dynamics~\cite{perez2019active,barton2017active}, and animal groups such as bird flocks~\cite{toner2024physics} to bacterial swarms~\cite{aranson2022bacterial}, self-rotating “spinners”~\cite{farhadi2018dynamics,van2016spatiotemporal}, and vibrated grains that behave as active particles at the tabletop scale~\cite{narayan2007long,ramaswamy2017active,gupta2022active}.

\vspace{8pt}
\noindent
Within this broad class, active Brownian particles (ABPs) provide a minimal model for self-propelled motion: a constant-magnitude propulsion along an internal orientation that itself undergoes stochastic dynamics, typically modeled with white noise for both center-of-mass and orientation degrees of freedom. Alternative formalisms-e.g., active Ornstein–Uhlenbeck particles or run-and-tumble dynamics-share equivalent low-order statistics with ABPs in many regimes~\cite{cates2013active,fodor2016far,das2018confined,solon2015active,shee2020active}. In numerous microscopic settings, the inertial relaxation time is much shorter than the orientation persistence time, rendering the overdamped approximation highly accurate.

\vspace{8pt}
\noindent
However, inertia cannot always be neglected~\cite{hecht2022active,te2023microscopic,sandoval2020pressure,nguyen2021active,de2022motility,lowen2020inertial,caprini2022role,omar2023tuning,chatterjee2021inertia,su2021inertia,caprini2021spatial,patel2023exact,mandal2019motility,herrera2021maxwell}. It becomes prominent in vibrobots and other granular realizations~\cite{deblais2018boundaries}, hexbug-like walkers~\cite{tapia2021trapped}, and fast active particles at higher Reynolds numbers~\cite{alert2022active}, where the relevant dynamical scales interpolate between Brownian and Fokker-Planck times~\cite{mayer2022inertial}. In such underdamped regimes (sometimes dubbed microflyers rather than microswimmers~\cite{lowen2020inertial}), inertia reshapes kinetics and transport. Experiments with symmetric granular rod robots show that inertia can suppress boundary accumulation and promote gas-like phases via local alignment hindrance~\cite{deblais2018boundaries}. More broadly, while passive long-time diffusion is insensitive to inertia, active particles with translational and rotational inertia display fundamentally different long-range transport set by the moment of inertia~\cite{scholz2018inertial}. When chirality coexists with inertia, additional phenomena appear-for instance, anisotropic vibrating rods cluster at boundaries, unlike isotropic round particles~\cite{kudrolli2008swarming}.

\vspace{8pt}
\noindent
Alongside inertia, chirality has emerged as a central organizing principle in active matter~\cite{sevilla2016diffusion,caprini2019active,levis2019activity,liebchen2022chiral,bickmann2022analytical,Caprini2023,debets2023glassy,ai2015chirality,caprini2024self,shee2024emergent}. Chirality can arise from intrinsic structural asymmetry or from dynamically generated rotation, and in both cases it biases trajectories and breaks time-reversal symmetry in ways not captured by achiral models~\cite{kummel2013circular,bechinger2016active,mano2017optimal,zhang2020reconfigurable,chan2024chiral,kaur2025novo,kant2025edge}. In the deterministic limit, chiral ABPs (cABPs) execute circular or helical paths in 2D or 3D, respectively, a kinematic hallmark seen across biology and engineered systems. Biological examples include intrinsically chiral starfish embryos that self-assemble into chiral crystal-like arrays~\cite{tan2022odd} and sperm cells whose chiral swimming constrains and optimizes search strategies in the reproductive tract~\cite{zaferani2023biphasic}. In synthetic and granular realizations, shape-asymmetric microswimmers exploit chirality for environmental sorting~\cite{mijalkov2013sorting}, and asymmetric ellipsoids display robust chiral motion~\cite{witten2020review}. These systems highlight chirality as a source of irreversibility beyond simple self-propulsion. 

\vspace{8pt}
\noindent
Chiral dynamics has been investigated in both two- and three-dimensional geometries within a wide variety of modeling frameworks~\cite{van2008dynamics,wittkowski2012self,kummel2013circular,volpe2014simulation,sevilla2016diffusion,lowen2016chirality,Kurzthaler2017a,chepizhko2019ideal,otte2021statistics,perez2019active,pattanayak2024impact,santra2025universal}.
Further, in “odd” active matter, chirality may emerge spontaneously giving rise to unconventional transport and mechanical responses—including odd viscosity, elasticity, and diffusivity~\cite{Soni2019,scheibner2020odd,Hargus2021a,fruchart2023odd,kole2021layered}. A growing body of experiments further shows that the transport properties of cABPs, such as effective diffusivity, are highly sensitive to the embedding medium and boundary conditions, in stark contrast to many achiral Brownian systems where coarse transport is comparatively robust~\cite{chan2024chiral}. At the collective level, strong chirality can disrupt motility-induced phase separation (MIPS), favoring dynamical clusters and altering steady-state structures~\cite{ma2022dynamical}. 

\vspace{8pt}
\noindent
A natural bridge between inertia and chirality is provided by tabletop ``vibrobot" realizations with Hexbugs, which let one dial chirality and probe finite-moment-of-inertia effects in a controlled way~\cite{barona2024playing}. Adding a small off-axis mass turns a nominal ABP into a chiral walker with steady angular velocity. Furthermore, the finite size of the robots and the momentum exchange with the substrate couple translation and rotation and can produce measurable lags between velocity and body orientation, especially after collisions.
Together, these bench-top observations underscore that, beyond overdamped limits, the joint action of chirality and inertia governs single-particle kinematics. 

\vspace{8pt}
\noindent
In this article, we explore the role of inertia in the dynamics of a cABP in two dimensions using a Laplace transform-based method which was originally introduced to study the dynamics of worm-like chain polymers~\cite{hermans1952statistics, daniels1952xxi} and has recently been adopted for active particle systems in various situations~\cite{shee2020active,chaudhuri2021active,shee2022active,patel2023exact,pattanayak2025chirality&trap,Patel2024,pattanayak2024impact}. We demonstrate exact analytical expressions of dynamical moments and show exact match with the numerical results.

\section{Langevin dynamics of a chiral active particle with inertia}
\begin{figure*}
    \centering
    \includegraphics[width=\linewidth]{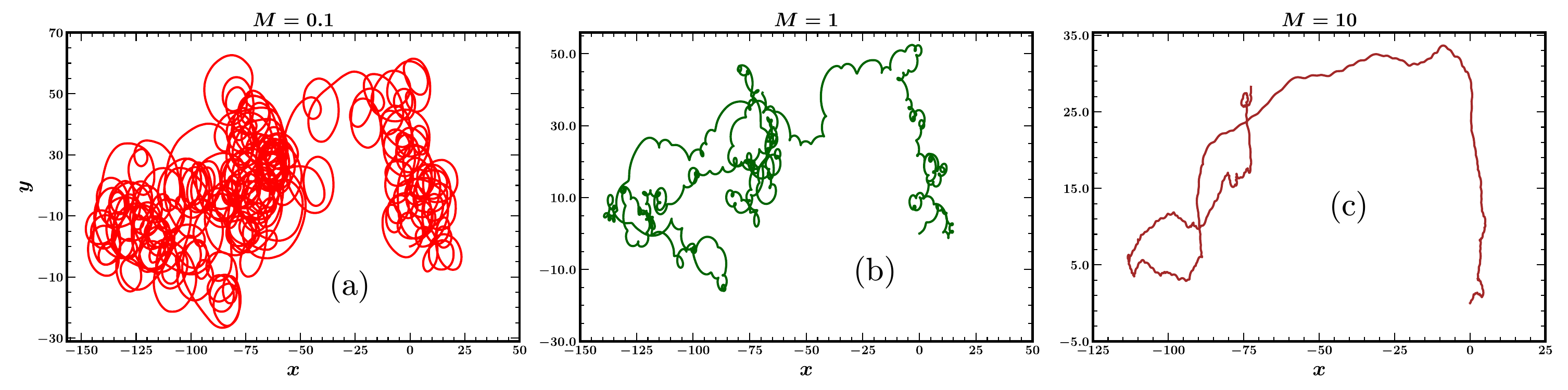}
    \caption{In this figure we have plotted the trajectory obtained by simulating the Langevin dynamics for fixed values of $\Pe$ and $\Omega$ and with different mass: (a) $M=0.1$, (b) $M=1$ and (c) $M=10$. For all figures (a)-(c) we have fixed $\Pe=100$ and $\Omega=100$.} 
    \label{trajectory_in}
\end{figure*}
\paragraph{Model.}
We consider a point-like chiral active Brownian particle (cABP) of finite mass $m$ moving in 2D with self-propulsion speed $v_0$ along the body orientation $\bu=(\cos\phi,\sin\phi)$, and an intrinsic (constant) angular velocity $\omega\hat{\mathbf z}$ normal to the plane.
The translational dynamics are \emph{underdamped}, while the orientation dynamics are \emph{overdamped}. In Cartesian components, the underdamped Langevin equations read (It\^o convention)
\begin{align}
     \begin{aligned}
     \dot{\bv} &= -\frac{\Gamma}{m} (\bv - v_0 \bu) +\frac{\Gamma}{m}\sqrt{2 D_t}\bm{\xi}(t)
     \\
     \dot{\br} &= \bv
     \\
     \dot{\phi} &= \omega + \sqrt{2 D_r}\zeta(t)
     \label{Langevin}
     \end{aligned}
\end{align}
where \(\Gamma\) is the Stokes drag, \(D_t\) and \(D_r\) are translational and rotational diffusivities, and \(\boldsymbol{\xi}(t)\) and \(\zeta(t)\) are mutually independent Gaussian white noises with zero mean and covariances
\(
\langle \xi_i(t)\xi_j(t')\rangle=\delta_{ij}\delta(t-t'),\quad
\langle \zeta(t)\zeta(t')\rangle=\delta(t-t').
\)

We can make the equations dimensionless using the rotational time scale $t_r=D_r^{-1}$ and the diffusive length scale $\ell=\sqrt{D_t/D_r}$. We define $\ttt = D_r t,\qquad \tbr={\bm r}/{\ell},\qquad
\tbv={\bm v}/{\sqrt{D_t D_r}}$ and introduce the dimensionless parameters
$M={mD_r}/{\Gamma}\quad\text{(dimensionless mass)},\qquad
\mathrm{Pe}={v_0}/{\sqrt{D_t D_r}}\quad\text{(active P\'eclet)},\qquad
\Omega={\omega}/{D_r}\quad\text{(dimensionless chirality)}$. 
Writing the noise in terms of independent Wiener processes  $\bm{B}(\ttt)$ and $W(\ttt)$ with
\(
\langle dB_i\,dB_j\rangle=\delta_{ij}\,d\ttt,\;
\langle dW^2\rangle=d\ttt,
\)
and $\bm{B}$ independent of $W$, the dimensionless stochastic differential equations (Eq.~\ref{Langevin}) become
\begin{align}
\begin{aligned}
d\tbv &= -\frac{1}{M}\Big(\tbv-\mathrm{Pe}\,\bu\Big)\,d\ttt + \frac{\sqrt{2}}{M}\,d\bm{B}(\ttt), \\
d\tbr &= \tbv\,d\ttt, \\
d\phi &= \Omega\,d\ttt + \sqrt{2}\,dW(\ttt). 
\label{Langevin_dimensionless}
\end{aligned}
\end{align}
Eq.~\eqref{Langevin_dimensionless} explicitly separate the \emph{underdamped} translational relaxation (controlled by \(M\)) from the \emph{overdamped} rotational diffusion with drift (controlled by \(\Omega\)). In the limits \(M\to 0\) and \(M\to\infty\) they respectively recover the overdamped ABP/cABP dynamics and the heavy-particle (inertial) regime.

\paragraph{Deterministic circular motion (noise-free).} \label{deterministic}
Setting $d\bm{B}=dW=0$ in Eqs.~\eqref{Langevin_dimensionless} gives
\begin{align}
    \dot{\tbv}=-\frac{1}{M}\Big(\tbv-\mathrm{Pe}\,\bu\Big), &&
\dot{\phi}=\Omega, && \dot{{\tbr}}={\tbv}
\label{Deterministic_eqn}
\end{align}
with $\bu=(\cos\phi,\sin\phi)$. Integrating Eq.~\ref{Deterministic_eqn} with respect to time, we obtain,
\begin{align*}
    \phi(\ttt) &= \Omega \ttt + \phi_0, 
    \\
    \tbv_x(\ttt) &= \tbv_{0x} e^{-\ttt/M}+\frac{\Pe}{M}e^{-\ttt/M}\int_0^\ttt e^{-\ttt^\prime/M} \cos{\left(\Omega \ttt^\prime+\phi_0\right)} d\ttt^\prime
    \\&=\tbv_{0x} e^{-\ttt/M} \\&+ \frac{ \Pe  \left(M \Omega  \sin (\Omega \ttt +\phi_0)+\cos ( \Omega \ttt+\phi_0)\right)-e^{-\ttt/M}\Pe (M \Omega  \sin (\phi_0)+\cos (\phi_0))}{M^2 \Omega ^2+1},
\end{align*}
\begin{align}
\begin{aligned}
    \tbv_y(\ttt) &= \tbv_{0y} e^{-\ttt/M}+\frac{\Pe}{M}e^{-\ttt/M}\int_0^\ttt e^{-\ttt^\prime/M} \sin{\left(\Omega \ttt^\prime+\phi_0\right)} d\ttt^\prime
    \\&=\tbv_{0y} e^{-\ttt/M} \\&+ \frac{ \Pe  \left(  \sin (\Omega \ttt +\phi_0)- M \Omega\cos ( \Omega \ttt+\phi_0)\right)-e^{-\ttt/M}\Pe (  \sin (\phi_0)-M\Omega\cos (\phi_0))}{M^2 \Omega ^2+1},
    \\
    {\tilde{\bm x}}(\ttt)& = \int_0^{\ttt} \tbv_x(\ttt^\prime) d\ttt^\prime= M \tbv_{0x}\left(1- e^{-\ttt/M}\right)+\frac{e^{-\frac{\ttt}{M}} \left(M \Pe \Omega  \left(M \Omega  \sin \left(\phi _0\right)+\cos \left(\phi _0\right)\right)\right)}{M^2 \Omega ^3+\Omega }
    \\ & 
    -\frac{\Pe \left(M \Omega\cos \left( \Omega \ttt +\phi _0\right)-\sin \left( \Omega \ttt+\phi _0\right)\right)}{M^2 \Omega ^3+\Omega }-\frac{\Pe \sin{\phi_0}}{\Omega},
    \\ 
    {\tilde{\bm y}}(\ttt)& = \int_0^\ttt \tbv_y(\ttt^\prime) d\ttt^\prime= M \tbv_{0y}\left(1- e^{-\ttt/M}\right)+ \frac{\Pe e^{-\frac{\ttt}{M}} \left(M \Omega  \left(\sin \left(\phi _0\right)-M \Omega  \cos \left(\phi _0\right)\right)\right)}{M^2 \Omega ^3+\Omega }
    \\& +\frac{\Pe \cos (\phi_0)}{\Omega } -\frac{\Pe \left(M \Omega  \sin \left( \Omega \ttt +\phi _0\right)+\cos \left( \Omega \ttt+\phi _0\right)\right)}{M^2 \Omega ^3+\Omega }. 
\end{aligned}
\end{align}
In the long-time limit,
\begin{align}
   \tbv_x(\ttt)^2+\tbv_y(\ttt)^2 &=\frac{\Pe^2}{1+M^2\Omega^2}
   \\
\left({\tilde{\bm x}}(\ttt)-a\right)^2 + \left({\tilde{\bm y}}(\ttt)-b\right)^2 &= \frac{\Pe^2}{\Omega^2(1+M^2\Omega^2)}
\end{align}
where $a= -\frac{\Pe \sin{\phi_0}}{\Omega}+ M \tbv_{0x}$ and $b= \frac{\Pe \cos{\phi_0}}{\Omega}+ M \tbv_{0y}$.
The magnitude of the velocity becomes $\lvert \tbv\rvert= \sqrt{\tbv_x(\ttt)^2+\tbv_y(\ttt)^2} =\frac{\Pe}{\sqrt{1+ M^2\Omega^2}}$ which reduces to $\Pe$ for both the overdamped ABP's($M\to 0$) and  achiral ABP's($\Omega\to0$). The position vector of the particle exhibits a circular trajectory with radius $\tilde{R}=\frac{\Pe}{\Omega\sqrt{1+ M^2 \Omega^2}}$. In the $ M\to 0$ limit, $\tilde R$ reduces to $\frac{\Pe}{\Omega}$, which is the radius of the noiseless trajectory of an overdamped chiral ABP. 
The circular trajectory is observed when the observation time is longer than the inertial time-scale, i.e. $\ttt> M_{in}$. During $\ttt\leq M$, we observe incomplete circles (see Fig.~\ref{fig:noiseless_trajectory}(b) in the appendix for reference). 

\subsection{Fokker-Planck equation and moment generating equation}
The probability density $P(\tbv,\tbr,\phi,\ttt)$ governed by Eqs.~\eqref{Langevin_dimensionless} (It\^o) satisfies the Fokker--Planck equation
 \begin{align}
   \nonumber \partial_\ttt P =& -\nabla_{\tbr}\cdot(\tbv P)+\frac{1}{M}\nabla_{\tbv}\cdot[(\tbv-\Pe\bu)P]+\frac{1}{M^2}\nabla_{\tbv}^2 P\\&-\partial_\phi (\Omega P)+\partial^2_\phi P
 \end{align}
 By performing a Laplace transform $\tilde{P}(\tbv,\tbr,\phi,s) = \int_0^\infty d\ttt e^{-s\ttt} P (\tbv,\tbr,\phi,\ttt)$, the Fokker-Planck equation can be expressed as
 \begin{align}
    \nonumber
     &s\tilde{P}(\tbv,\tbr,\phi,s)= -P(\tbv,\tbr,\bu,0)-\nabla_{\tbr}\cdot(\tbv \tilde{P})\\&+\frac{1}{M}\nabla_{\tbv}\cdot[(\tbv-\Pe\bu)\tilde{P}]+\frac{1}{M^2}\nabla_{\tbv}^2 \tilde{P}-\partial_\phi (\Omega \tilde{P})+\partial^2_\phi \tilde{P}
     \label{eq:FP_Laplace}
 \end{align}
 The initial condition is set by $P(\tbv,\tbr,\phi,0)=\delta(\tbv-\tbv_0)\delta(\tbr)\delta(\phi -\phi_0)$. We can finally obtain the moment-generating function by considering an arbitrary function $\psi=\psi(\tbv,\tbr,\phi)$. Multiplying Eq.~\eqref{eq:FP_Laplace} by $\psi$ and integrating over all variables, with integrations by parts, yields the Laplace-domain \emph{moment identity}:
 \begin{align}
 \begin{aligned}
     & s \langle \psi\rangle_s= \langle\psi\rangle_0+ \langle \tbv\cdot\nabla_\tbr \psi\rangle+\frac{1}{M}\langle (\Pe~\bu-\tbv)\cdot\nabla_\tbv \psi\rangle\\&+\frac{1}{M^2}\langle\nabla^2_\tbv \psi\rangle+\langle\partial_\phi^2\psi\rangle_s+\Omega\langle\partial_\phi\psi\rangle_s
     \label{ME_inertia}
\end{aligned}
 \end{align}
 where $\langle\psi\rangle_s = \int d\tbv d\tbr d\phi \psi(\tbv,\tbr,\phi)\tilde{P}(\tbv,\tbr,\phi,s)$ and $\langle \psi \rangle_0 = \int d\tbv d\tbr d\phi\, \psi(\tbv,\tbr,\phi)\, P(\tbv,\tbr,\phi,0)$.
 Equation~\eqref{ME_inertia} generates a closed hierarchy of Laplace-domain moment equations by suitable choices of $\psi$. Inverse Laplace transforms then yield the full time-dependent moments. Note that $\langle \cdot \rangle$ denote ensemble averages while $\langle \cdot \rangle_s$ denote the Laplace transform average. 
 
 \section{Calculation of different moments}

 \subsection{Orientation auto-correlation}
 We first consider $\psi = \bu\cdot\bu_0$ where $\bu=(\cos\phi,\sin\phi)$ and $\bu_0=(\cos\phi_0,\sin\phi_0)$ denotes the initial orientation.
 Repeating the procedure described above, we get the time-dependent expression of the orientation auto-correlation function,
 \begin{align}
    \langle\bu\cdot\bu_0\rangle=e^{- \ttt}\cos (\Omega \ttt)\,.
\label{orientation_autocorr_inertia} 
 \end{align}
 As expected, the orientation autocorrelation is \emph{independent} of the translational parameters $M$ and $\Pe$, because the rotational sector is overdamped and autonomous in the present model.
 
 \section{Mean velocity}

We now consider $\psi=\tbv$. From the Laplace-domain moment identity in Eq.~\eqref{ME_inertia},
choosing $\psi=\tbv$ gives the vector equation
\begin{equation}
\Big(s+\frac{1}{M}\Big)\,\langle\tbv\rangle_s
= \langle\tbv\rangle_{0} + \frac{\Pe}{M}\,\langle\bu\rangle_s.
\label{eq:mean_v_Laplace}
\end{equation}
The time-dependent expressions of the mean velocity are given as,
\begin{align}
\begin{aligned}
\big\langle\tbv_x(\ttt)\big\rangle
&= \tbv_{0x}\,e^{-\ttt/M}
+ \frac{\Pe}{M^2\Omega^2+(M-1)^2}\Big\{
e^{-\ttt/M}\big[(M-1)\cos\phi_0 - M\Omega\sin\phi_0\big]
\\
&\hspace{7.8em}
+ e^{-\ttt}\big[M\Omega\sin(\Omega\ttt+\phi_0) - (M-1)\cos(\Omega\ttt+\phi_0)\big]
\Big\},
\\[4pt]
\big\langle\tbv_y(\ttt)\big\rangle
&= \tbv_{0y}\,e^{-\ttt/M}
+ \frac{\Pe}{M^2\Omega^2+(M-1)^2}\Big\{
e^{-\ttt/M}\big[(M-1)\sin\phi_0 + M\Omega\cos\phi_0\big]
\\
&\hspace{7.8em}
- e^{-\ttt}\big[(M-1)\sin(\Omega\ttt+\phi_0) + M\Omega\cos(\Omega\ttt+\phi_0)\big]
\Big\}.
\end{aligned}
\label{first_moment_velocity}
\end{align}
Both components vanish in the long-time limit, $\langle\tbv_x\rangle,\langle\tbv_y\rangle\to 0$

\begin{figure*}
\centering
\includegraphics[width=0.7\linewidth]{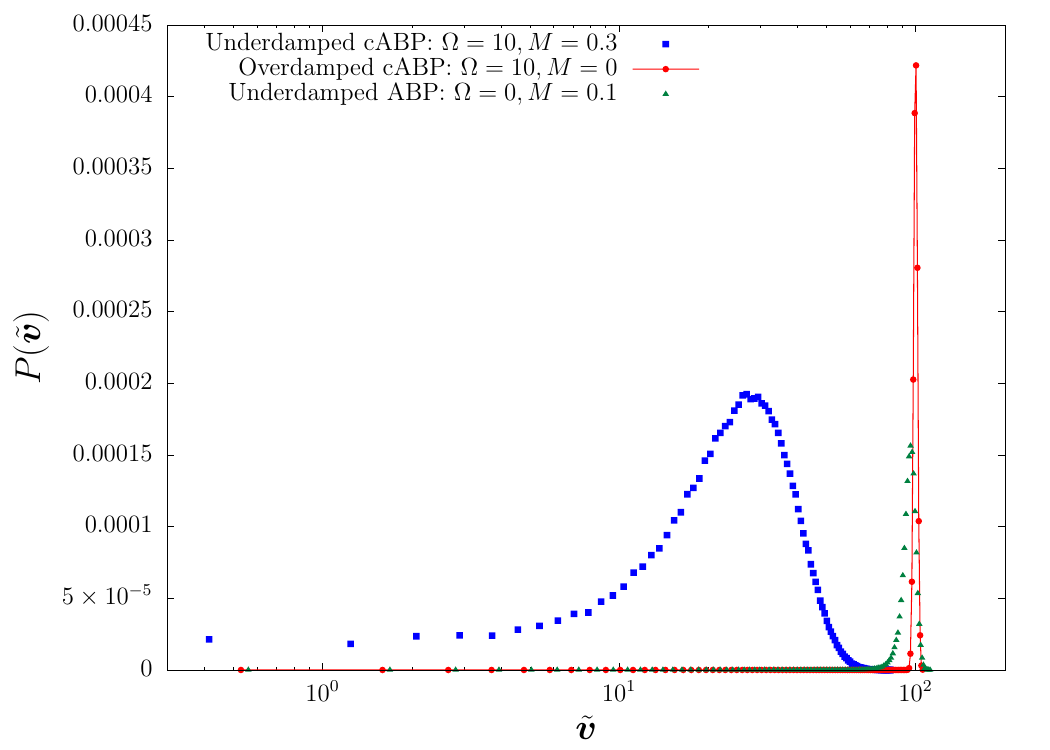}
\caption{Probability density of the speed, $|\tbv|=\sqrt{\tbv_x^2+\tbv_y^2}$, from simulations of $N$ particles.
For the overdamped cABP and the underdamped achiral ABP, the peak is at $|\tbv|\approx\Pe=100$.
For an underdamped \emph{chiral} ABP, inertia and chirality shift the peak to
$\tbv_m \approx \Pe/\sqrt{1+M^2\Omega^2}$.}
\label{prob_abs_vel}
\end{figure*}

\paragraph{Noiseless steady (periodic) state.}
In the deterministic limit (no translational/rotational noise), the long-time response is a periodic orbit
at frequency $\Omega$ with components
\begin{align}
\begin{aligned}
\big\langle\tbv_x(\ttt)\big\rangle^{\text{st}}_{\text{nl}}
&= \frac{\Pe}{1+M^2\Omega^2}\Big[\sin(\Omega\ttt+\phi_0)-M\Omega\cos(\Omega\ttt+\phi_0)\Big],\\
\big\langle\tbv_y(\ttt)\big\rangle^{\text{st}}_{\text{nl}}
&= \frac{\Pe}{1+M^2\Omega^2}\Big[\cos(\Omega\ttt+\phi_0)+M\Omega\sin(\Omega\ttt+\phi_0)\Big],
\end{aligned}
\end{align}
so that the instantaneous amplitude is
\begin{equation}
\big|\langle\tbv(\ttt)\rangle\big|^{\text{st}}_{\text{nl}}
= \frac{\Pe}{\sqrt{\,1+M^2\Omega^2\,}}.
\end{equation}
For an overdamped cABP and for an inertial ABP without chirality ($\Omega=0$), the noiseless average speed equals $\Pe$. In Fig.~\ref{prob_abs_vel}, we have plotted the probability distribution of the magnitude of the velocity in the steady state obtained by simulating multiple particles. For inertial cABP, the peak of the distribution changes to $\frac{\Pe}{\sqrt{1+M^2\Omega^2}}$ from $\Pe$ for both overdamped cABP and inertial ABP.

\subsection{Velocity components parallel and perpendicular to the heading direction}
 The velocity along the direction of the heading direction of the active force can be written as $\tbv_{\parallel}= (\tbv\cdot\bu)\bu$. Setting setting $\psi=\tbv\cdot\bu$ in eq.~[\ref{ME_inertia}], we obtain $\left(s+1+\frac{1}{M}\right)\langle\tbv\cdot\bu\rangle_s=\tbv_0\cdot\bu_0 +\frac{\Pe}{M s} +\Omega \langle \partial_\phi(\tbv\cdot\bu)\rangle_s$. The term, $\langle \partial_\phi\tbv\cdot\bu\rangle_s$ can also be expressed as $\langle \tbv\cdot\bu^{\perp}\rangle_s$ where $\bu^{\perp}=-\sin{\phi}\hat{x}+\cos{\phi}\hat{y}$. To compute this, we set $\psi=\partial_\phi\tbv\cdot\bu$. The moment equation gives $\left(s+1+\frac{1}{M}\right)\langle\partial_\phi\tbv\cdot\bu\rangle_s=\tbv_0\cdot\bu^\perp_0 -\Omega \langle\tbv\cdot\bu\rangle_s$. Hence, on substitution, we get,
\begin{align}
\Big[s+1+\tfrac{1}{M}+\frac{\Omega^2}{s+1+\tfrac{1}{M}}\Big]\,
\langle\tbv\!\cdot\!\bu\rangle_s
= \tbv_0\!\cdot\!\bu_0
+ \frac{\Omega}{s+1+\tfrac{1}{M}}\,\tbv_0\!\cdot\!\bu_0^\perp
+ \frac{\Pe}{Ms}.
\end{align}
Inverting the Laplace transform yields the exact time dependence
\begin{align}
\begin{aligned}
\langle \tilde v_\parallel\rangle(\ttt)
&\equiv \langle\tbv\!\cdot\!\bu\rangle(\ttt)
= \frac{(M+1)\,\Pe}{(M+1)^2+M^2\Omega^2}
+ e^{-\frac{(M+1)\ttt}{M}}
\Big[
-\frac{(M+1)\,\Pe}{(M+1)^2+M^2\Omega^2}
+ \tbv_0\!\cdot\!\bu_0 - \Omega\,\tbv_0\!\cdot\!\bu_0^\perp
\Big]\cos(\Omega\ttt)
\\
&\quad
+ e^{-\frac{(M+1)\ttt}{M}}
\frac{M\,\Pe\,\Omega}{(M+1)^2+M^2\Omega^2}\,\sin(\Omega\ttt),
\end{aligned}
\label{vu}
\end{align}
while the perpendicular projection $\langle\tbv\!\cdot\!\bu^\perp\rangle(\ttt)$ is
\begin{align}
\begin{aligned}
\langle\tbv\!\cdot\!\bu^\perp\rangle(\ttt)
&= -\frac{M\,\Pe\,\Omega}{(M+1)^2+M^2\Omega^2}
+ e^{-\frac{(M+1)\ttt}{M}}\,\tbv_0\!\cdot\!\bu_0^\perp
+ e^{-\frac{(M+1)\ttt}{M}}
\frac{M\,\Pe\,\Omega}{(M+1)^2+M^2\Omega^2}\,\cos(\Omega\ttt)
\\
&\quad
+ e^{-\frac{(M+1)\ttt}{M}}
\frac{(M+1)\,\Pe - \big((M+1)^2+M^2\Omega^2\big)
\big(\tbv_0\!\cdot\!\bu_0 - \Omega\,\tbv_0\!\cdot\!\bu_0^\perp\big)}{(M+1)^2+M^2\Omega^2}\,
\sin(\Omega\ttt).
\end{aligned}
\end{align}
In the long-time limit,
\begin{align}
\lim_{\ttt\to\infty}\langle\tbv\!\cdot\!\bu\rangle(\ttt)
&= \frac{(M+1)\,\Pe}{(M+1)^2+M^2\Omega^2},
\\
\lim_{\ttt\to\infty}\langle\tbv\!\cdot\!\bu^\perp\rangle(\ttt)
&= -\frac{M\,\Omega\,\Pe}{(M+1)^2+M^2\Omega^2}.
\end{align}
Unlike the achiral inertial ABP~\cite{patel2023exact}, the \emph{perpendicular} projection does not vanish
as $\ttt\to\infty$ in the chiral case ($\Omega\neq 0$). In the overdamped ($M\to0$) limit, for both chiral and achiral particles, the perpendicular component vanishes and the parallel component approaches $\Pe$, indicating that the velocity of the overdamped ABP (chiral and achiral) always aligns with the orientation of the active force. 
\begin{figure*}
    \centering
    \includegraphics[width=\linewidth]{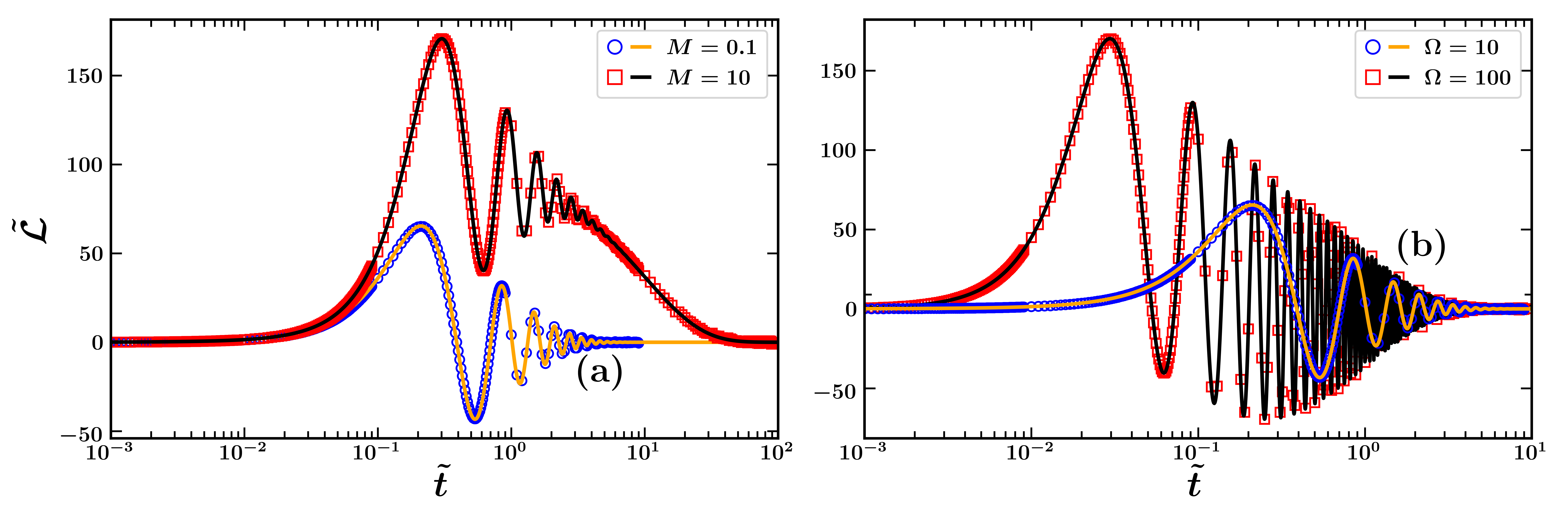}
    \caption{Lag function is plotted as a function of time for (a) fixed $\Pe=100$, $\Omega=10 $ and different values of effective mass ($M$) and for (b) fixed $\Pe=100$, $M=0.1$ and different chirality. We set the initial velocity $\tbv= 100\hat{x}$.}
    \label{Lag}
\end{figure*}
\subsubsection{Inertial delay/ lag}
The Langevin equations imply that $\langle\tbv\rangle$ need not be colinear with $\bu$ at a given time.
To quantify the time-asymmetric coupling between velocity and orientation, define the lag
\begin{align}
\begin{aligned}
\mathcal{L}(\ttt)=&\big\langle \tbv(\ttt)\!\cdot\!\bu(0)\big\rangle
- \big\langle \tbv(0)\!\cdot\!\bu(\ttt)\big\rangle
\\=& e^{-\frac{\ttt}{M}}\left(\langle \tbv(0)\cdot\bu(0)\rangle+\frac{\Pe(M-1)}{M^2\Omega^2+(M-1)^2}\right)
+\frac{\Pe}{M^2\Omega^2+(M-1)^2}e^{-\ttt}\left(M \Omega \sin{\Omega \ttt}-(M-1)\cos{\Omega \ttt}\right)\\&-e^{-t}\left(\cos{\Omega\ttt}\langle \tbv(0)\cdot\bu(0)\rangle+\sin{\Omega\ttt}\langle \tbv(0)\cdot\bu^\perp(0)\rangle\right)
\end{aligned}
\end{align}
As discussed in Ref.~\cite{lowen2020inertial}, $\mathcal{L}(0)=0$, it grows from zero,
attains a maximum at an intermediate time set by $M$ and $\Omega$, and then decays
back to zero as $\ttt\to\infty$, capturing the \emph{inertial delay} between orientation
and velocity. 
The effects of mass and chirality ($M,\Omega$) on lag are shown in Fig.~\ref{Lag}. As expected, $\mathcal{L}$ decreases as $M \rightarrow 0$ (Fig.~\ref{Lag}(a)).

\section{Mean displacement}

With $\psi=\tbr$ the Laplace-domain identity gives $\langle\tbr\rangle_s=\langle\tbv\rangle_s/s$.
Inverting the transform and using Eq.~\eqref{first_moment_velocity}, the time-dependent mean displacement
components read
\begin{align}
\begin{aligned}
\big\langle \tilde{x}(\ttt)\big\rangle
&= \frac{1}{(\Omega^2+1)\,\big(M(M\Omega^2+M-2)+1\big)}\Big[
 - M(\Omega^2+1)\,e^{-\ttt/M}\Big(-M\,\Pe\,\Omega\,\sin\phi_0
\\[-2pt]
&\qquad
+ (M-1)\Pe\,\cos\phi_0 + M\tbv_{0x}\,(M\Omega^2+M-2) + \tbv_{0x}\Big)
\\
&\qquad
- \Pe\,e^{-\ttt}\Big(\big(M(\Omega^2-1)+1\big)\cos(\Omega\ttt+\phi_0)
+ (2M-1)\Omega\,\sin(\Omega\ttt+\phi_0)\Big)
\\
&\qquad
+ \big(M(M\Omega^2+M-2)+1\big)\Big(M\tbv_{0x}(\Omega^2+1)
 - \Pe\,\Omega\,\sin\phi_0 + \Pe\,\cos\phi_0\Big)\Big],
\\[6pt]
\big\langle \tilde{y}(\ttt)\big\rangle
&= \frac{1}{(\Omega^2+1)\,\big(M(M\Omega^2+M-2)+1\big)}\Big[
 - M(\Omega^2+1)\,e^{-\ttt/M}\Big(M\,\Pe\,\Omega\,\cos\phi_0
\\[-2pt]
&\qquad
+ (M-1)\Pe\,\sin\phi_0 + M\tbv_{0y}\,(M\Omega^2+M-2) + \tbv_{0y}\Big)
\\
&\qquad
- \Pe\,e^{-\ttt}\Big(\big(M(\Omega^2-1)+1\big)\sin(\Omega\ttt+\phi_0)
+ (1-2M)\Omega\,\cos(\Omega\ttt+\phi_0)\Big)
\\
&\qquad
+ \big(M(M\Omega^2+M-2)+1\big)\Big(M\tbv_{0y}(\Omega^2+1)
 + \Pe\,\Omega\,\cos\phi_0 + \Pe\,\sin\phi_0\Big)\Big].
\end{aligned}
\label{eq:mean_r_components}
\end{align}

In the long-time limit, the mean displacement approaches
\begin{align}
\begin{aligned}
\big\langle \tilde{x} \big\rangle\Big|_{\ttt\to\infty}
&= M\,\tbv_{0x} + \frac{\Pe}{\Omega^2+1}\,\big(\cos\phi_0 - \Omega \sin\phi_0\big),\\
\big\langle \tilde{y} \big\rangle\Big|_{\ttt\to\infty}
&= M\,\tbv_{0y} + \frac{\Pe}{\Omega^2+1}\,\big(\sin\phi_0 + \Omega \cos\phi_0\big).
\end{aligned}
\label{eq:mean_r_longtime}
\end{align}

\paragraph{Interpretation and timescales.}
The structure of Eq.~\eqref{eq:mean_r_components} cleanly separates two physical
contributions:
(i) an \emph{inertial} transient $\propto e^{-\ttt/M}$ stemming from the underdamped
translation, and (ii) a \emph{chiral–rotational} contribution $\propto e^{-\ttt}\{\cos(\Omega\ttt+\phi_0),\sin(\Omega\ttt+\phi_0)\}$ from the overdamped
orientation with drift~$\Omega$. Three characteristic timescales control the transient
kinematics:
\[
\tau_{\rm in}=M,\qquad \tau_{\rm rot}=1,\qquad \tau_{\rm circ}=\Omega^{-1}.
\]
For $M\ll 1$ (overdamped translation), $e^{-\ttt/M}$ decays rapidly and the
damped-oscillatory chiral term dominates; pronounced circular segments are visible whenever
$\Omega\gtrsim 1$ (i.e.\ $\tau_{\rm circ}\lesssim\tau_{\rm rot}$).
For $M\gg 1$ (strong inertia), the $e^{-\ttt/M}$ contribution becomes dominant.
\paragraph{Noiseless limit.}
In the deterministic case ($D_t=D_r=0$), the $e^{-\ttt}$ term is absent and the trajectory
is a circle of radius $R=\Pe/(\Omega\sqrt{1+M^2\Omega^2})$ (Sec. ~\ref{deterministic}), but the \emph{visibility} of circular motion is controlled by the separation of scales:
for $\ttt\lesssim\tau_{\rm in}$ the motion is inertia-dominated (little rotation visible),
while for $\ttt\gtrsim\tau_{\rm in}$ the rotation is manifest. 

\begin{figure*}[h]
    \includegraphics[width=\linewidth]{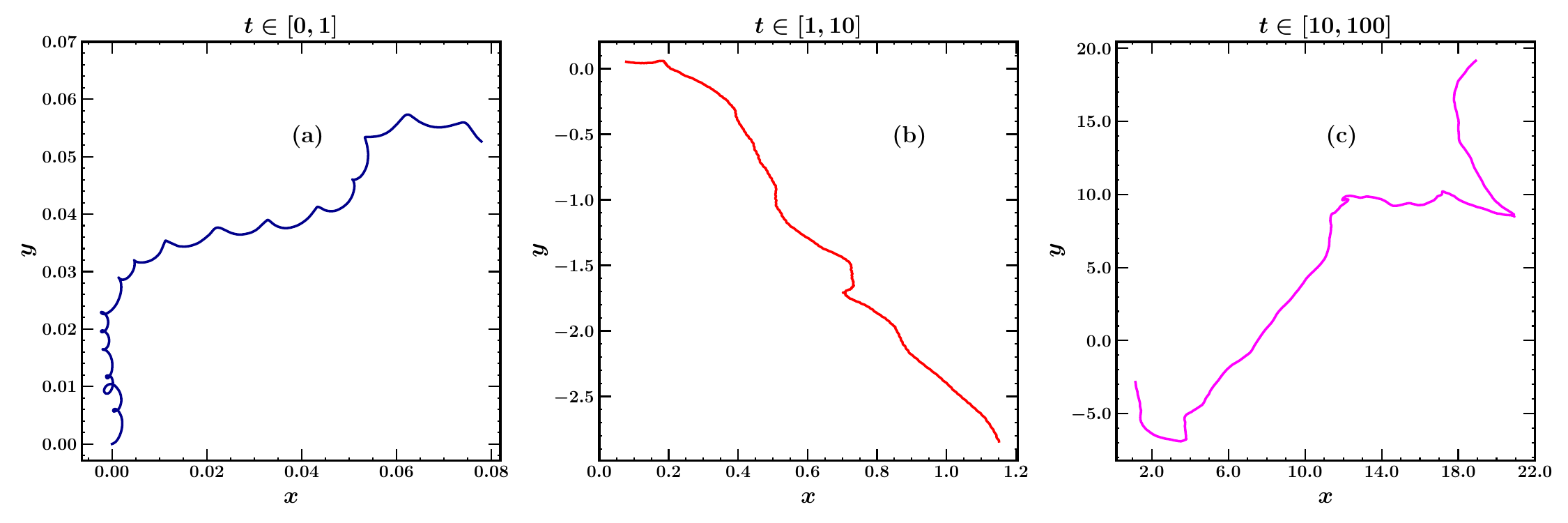}
    \caption{Trajectories at different time intervals for $M=10$, $\Pe=100$, $\Omega=100$. The rotational nature gradually decays as we keep on increasing time.}
    \label{trajectory_mass10_time}
\end{figure*}

\vspace{8pt}
\noindent
From the noiseless trajectory, we can conclude that the circular trajectory does not appear as long as $e^{-\frac{\ttt}{M}}$ does not decay to zero. Therefore, when $M>1$, the inertial term $e^{-\frac{\ttt}{M}}$ dominates over the decaying oscillatory term, $e^{-\ttt}\{\cos(\Omega\ttt+\phi_0),\sin(\Omega\ttt+\phi_0)\}$, which means that the full rotation of the orientation angle can not take place as it gets reoriented because of rotational diffusion. This explains the loss of apparent rotational structure with increasing~$M$
in Fig.~\ref{trajectory_in} at fixed $(\Omega,\Pe)$. The presence of $M$ in the denominator of the oscillatory terms is also responsible for the invisibility of circular traces in the trajectory. In Fig.~\ref{trajectory_mass10_time}, we have shown how the trajectory evolves with time and how the incomplete circular traces become invisible as we observe longer and the length-scale becomes larger.

\paragraph{Nonzero mean displacement from initial velocity.}
Eq.~\eqref{eq:mean_r_longtime} highlights the inertial ``memory’’ of the initial
velocity: even though for long-time limit $\langle\tbv(\ttt)\rangle\to 0$, the integrated inertial
transient yields a finite drift $M\,\tbv_0$ in the mean displacement. Thus any
nonzero $\tbv_0$ produces a \emph{nonzero} asymptotic $\langle\tbr\rangle$, on top of
the chiral contribution $\Pe(\bu_0+\Omega\bu_0^\perp)/(\Omega^2+1)$.

  \begin{figure*}
    \centering
    \includegraphics[width=\linewidth]{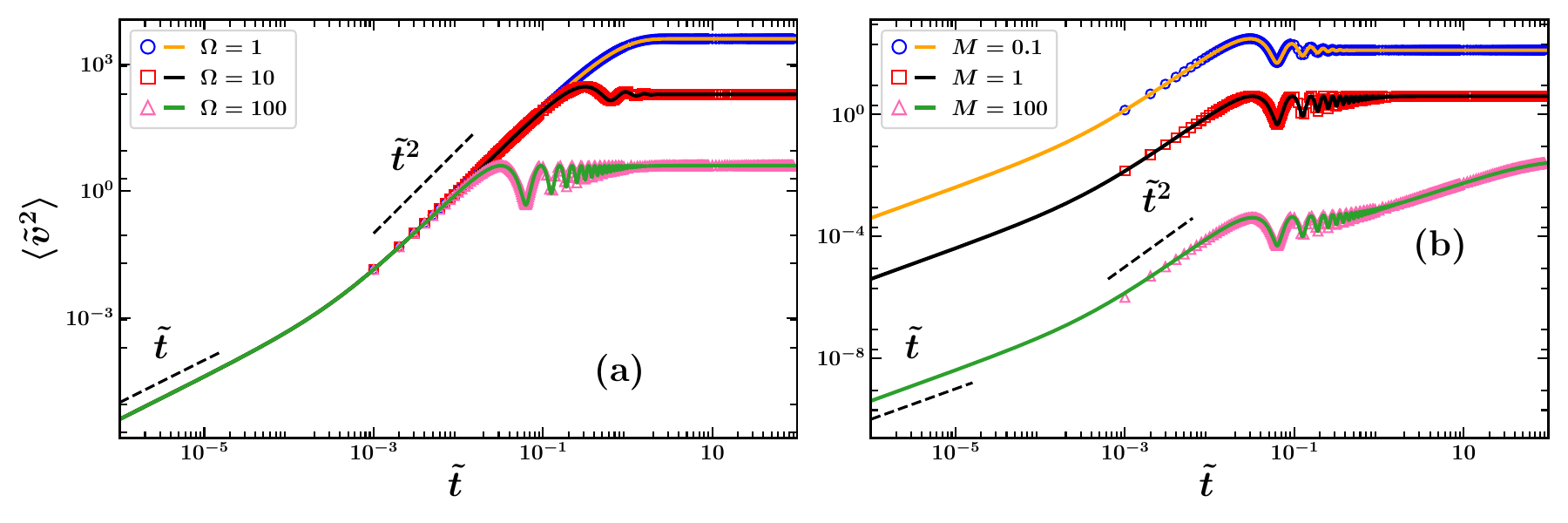}
    \caption{Mean squared velocity (MSV) is plotted against for (a) various $\Omega$ and a fixed $M=1$ and (b) various mass $M$ and a fixed $\Omega=100$. We have fixed $\Pe=100$ for both plots. From both (a) and (b) we can conclude that MSV eventually goes to its steady state in the long time limit. In the limit of $t<<1$, MSV first scales as $~\ttt$ and then scales as $\ttt^2$ in the intermediate time scale. The higher the mass the longer MSV takes to reach the steady state and the steady state value of MSV also gets lower. The lines are plotted from the exact analytical expression given in Eq.~\eqref{MSV_ana_in} and points are obtained by simulating the Langevin equations, Eq.~\eqref{Langevin_dimensionless}.}
    \label{msv_in}
\end{figure*}

\section{Mean squared velocity}
\label{MSV_in}

Choosing $\psi=\tbv^2$ in Eq.~\eqref{ME_inertia} gives
\begin{equation}
\Big(s+\frac{2}{M}\Big)\,\langle \tbv^2\rangle_s
= \tbv_0^2 + \frac{2\Pe}{M}\,\langle\tbv\!\cdot\!\bu\rangle_s
+ \frac{4}{M^2 s}.
\label{eq:MSV_Laplace}
\end{equation}
To close the system we also need $\langle\tbv\!\cdot\!\bu\rangle_s$ and
$\langle\tbv\!\cdot\!\bu^\perp\rangle_s$. Using $\psi=\tbv\!\cdot\!\bu$ and
$\psi=\tbv\!\cdot\!\bu^\perp$ in Eq.~\eqref{ME_inertia} yields
\begin{align}
\Big(s+1+\tfrac{1}{M}\Big)\,\langle\tbv\!\cdot\!\bu\rangle_s
&= \tbv_0\!\cdot\!\bu_0 + \frac{\Pe}{Ms}
+ \Omega\,\langle\tbv\!\cdot\!\bu^\perp\rangle_s,
\\
\Big(s+1+\tfrac{1}{M}\Big)\,\langle\tbv\!\cdot\!\bu^\perp\rangle_s
&= \tbv_0\!\cdot\!\bu_0^\perp - \Omega\,\langle\tbv\!\cdot\!\bu\rangle_s,
\end{align}
whose elimination gives
\begin{equation}
\Big[s+1+\tfrac{1}{M}+\frac{\Omega^2}{\,s+1+\tfrac{1}{M}\,}\Big]\,
\langle\tbv\!\cdot\!\bu\rangle_s
= \tbv_0\!\cdot\!\bu_0
+ \frac{\Omega}{\,s+1+\tfrac{1}{M}\,}\,\tbv_0\!\cdot\!\bu_0^\perp
+ \frac{\Pe}{Ms}.
\label{eq:v_dot_u_closed}
\end{equation}
Substituting Eq.~\eqref{eq:v_dot_u_closed} into Eq.~\eqref{eq:MSV_Laplace} and inverting the Laplace
transform yields the exact time dependence:
\begin{align}
\begin{aligned}
\langle \tbv^2(\ttt)\rangle
&= \frac{e^{-2\ttt/M}}{M^5(\Omega^2+1)^2+2M^3(\Omega^2-1)+M}\,\Bigg[
-2M\Pe^2 e^{(\frac{1}{M}-1)\ttt}\Big(\big(M^2(\Omega^2-1)+1\big)\cos(\Omega\ttt)
\\
&\qquad\qquad\qquad\qquad\qquad\qquad\quad
+ 2M^2\Omega\,\sin(\Omega\ttt)\Big)
+ \big(M(M\Omega^2+M+2)+1\big)
\\
&\quad\times\Big(M^2\Omega^2(M\tbv_0^2-2) + (M-1)\big(-M(\Pe^2+2)+(M-1)M\tbv_0^2+2\big)\Big)
\\
&\quad + e^{2\ttt/M}\big(M(M\Omega^2+M-2)+1\big)\Big(M\big(M(\Pe^2+2\Omega^2+2)+\Pe^2+4\big)+2\Big)
\\
&\quad + 2M\Pe\,e^{-\ttt}\big(M(M\Omega^2+M+2)+1\big)\Big(
e^{\ttt}\big[\tbv_{0x}((M-1)\cos\phi_0 - M\Omega\sin\phi_0)
\\
&\quad
+ \tbv_{0y}(M\Omega\cos\phi_0 + (M-1)\sin\phi_0)\big]
+ e^{\ttt/M}\big[\tbv_{0x}(M\Omega\sin(\Omega\ttt+\phi_0)
\\
&\quad-(M-1)\cos(\Omega\ttt+\phi_0)) + \tbv_{0y}(-(M-1)\sin(\Omega\ttt+\phi_0)
- M\Omega\cos(\Omega\ttt+\phi_0))\big]\Big)
\Bigg].
\end{aligned}
\label{MSV_ana_in}
\end{align}
We have plotted $\langle\tbv^2(\ttt)\rangle$ as a function of time in Fig.~\ref{msv_in}, illustrating the effects of chirality and inertia on the transient and steady-state behaviors of the mean squared velocity. We now discuss the various limits. 

\paragraph{Steady state.}
In the long-time limit ($\ttt\to\infty$), the mean-squared velocity approaches
\begin{equation}
{\;
\langle \tbv^2\rangle_{\mathrm{st}}
= \frac{(M+1)\,\Pe^2}{\,M(M\Omega^2+M+2)+1\,} + \frac{2}{M}\,.
\;}
\label{eq:MSV_steady}
\end{equation}
This dependence of the saturation value on $(\Omega,M)$ is seen in Fig.~\ref{msv_in}. 

\paragraph{Short-time expansion.}
For $\ttt\to 0$, we obtain
\begin{align}
\begin{aligned}
\langle \tbv^2(\ttt)\rangle
&= \tbv_0^2
+ \ttt\,\frac{2M\Pe\,\tbv_0\!\cdot\!\bu_0 - 2M\tbv_0^2 + 4}{M^2}
\\
&\quad
+ \ttt^2\,\frac{M\Pe\big(- (M+3)\tbv_0\!\cdot\!\bu_0 + M\Omega\,\tbv_0\!\cdot\!\bu_0^\perp + \Pe\big)
+ 2M\tbv_0^2 - 4}{M^3}
+ \mathcal{O}(\ttt^3).
\end{aligned}
\end{align}
As is evident from the expression above, the initial $\ttt$ dependent part is independent of $\Omega$. This is clearly seen in Fig.~\ref{msv_in}(a). As the $\ttt^2$ dependence sets in, $\tbv^2(\ttt)$ starts to separate. At a fixed $\Omega$, the variation with $M$ is clearly distinct at all $\ttt$ (Fig.~\ref{msv_in}(b)).  

\paragraph{Limiting behaviors of $\langle \tbv^2\rangle_{\mathrm{st}}$.} We now list the limiting values of $\langle \tbv^2\rangle_{\mathrm{st}}$ up to the second order.
\begin{align}
\lim_{M\to 0}\langle \tbv^2\rangle_{\mathrm{st}}
&= \frac{2}{M} + \Pe^2 - M\Pe^2 + \mathcal{O}(M^2),\\
\lim_{M\to \infty}\langle \tbv^2\rangle_{\mathrm{st}}
&= \Big(\frac{\Pe^2}{\Omega^2+1}+2\Big)\frac{1}{M}
+ \frac{\Pe^2(\Omega^2-1)}{M^2(\Omega^2+1)^2} + \mathcal{O}(M^{-3}),
\label{vst_heavy_mass}\\
\lim_{\Omega\to 0}\langle \tbv^2\rangle_{\mathrm{st}}
&= \frac{\Pe^2}{M+1} + \frac{2}{M}
 - \frac{M^2\Pe^2 \Omega^2}{(M+1)^3} + \mathcal{O}(\Omega^3),\\
\lim_{\Omega\to \infty}\langle \tbv^2\rangle_{\mathrm{st}}
&= \frac{2}{M} + \frac{(M+1)\Pe^2}{M^2\Omega^2} + \mathcal{O}(\Omega^{-3}).
\end{align}

\subsection{Components of MSV}

Setting $\psi=\tbv_x^2$, $\tbv_y^2$, and $\tbv_x\tbv_y$ yields the Laplace-domain components
(for brevity we keep $\tbv_0$ arbitrary here; later we set $\tbv_0=\bm 0$):
\begin{align}
\begin{aligned}
\langle \tbv_x^2\rangle_s
&= \frac{2}{Ms(2+Ms)}
\\
&\quad
+ 2\Pe^2\,
\frac{(s+4)(Ms+M+1)(s\,\bu_{x0}^2+2) - s\,\bu_{x0}\bu_{y0}\,\Omega\,(3M(s+2)+2)}
{s(Ms+2)\big((s+4)^2+4\Omega^2\big)\big(M^2\Omega^2+(Ms+M+1)^2\big)}
\\
&\quad
+ 2\Pe^2\,
\frac{\Omega^2\Big(M\big(s(3-2\bu_{x0}^2)+2\big)+2\Big)}
{s(Ms+2)\big((s+4)^2+4\Omega^2\big)\big(M^2\Omega^2+(Ms+M+1)^2\big)},
\\[4pt]
\langle \tbv_y^2\rangle_s
&= \frac{2}{Ms(2+Ms)}
\\
&\quad
+ 2\Pe^2\,
\frac{s\,\bu_{x0}\bu_{y0}\,\Omega\,(3M(s+2)+2) + (s+4)(Ms+M+1)(s\,\bu_{y0}^2+2)}
{s(Ms+2)\big((s+4)^2+4\Omega^2\big)\big(M^2\Omega^2+(Ms+M+1)^2\big)}
\\
&\quad
+ 2\Pe^2\,
\frac{\Omega^2\Big(M\big(2s\,\bu_{x0}^2 + s + 2\big)+2\Big)}
{s(Ms+2)\big((s+4)^2+4\Omega^2\big)\big(M^2\Omega^2+(Ms+M+1)^2\big)},
\\[4pt]
\langle \tbv_x\tbv_y\rangle_s
&= \Pe^2\,
\frac{\big(2\bu_{x0}^2-1\big)\Omega\,(3M(s+2)+2)
+ 2(s+4)\bu_{x0}\bu_{y0}\,(Ms+M+1)
- 4M\bu_{x0}\bu_{y0}\,\Omega^2}
{(Ms+2)\big((s+4)^2+4\Omega^2\big)\big(M^2\Omega^2+(Ms+M+1)^2\big)}.
\end{aligned}
\end{align}
The time-dependent components follow by inverse Laplace transform. Averaging over a uniform
initial orientation $\phi_0\in[0,2\pi]$ yields $\langle\tbv_x\tbv_y\rangle=0$ and
$\langle\tbv_x^2\rangle=\langle\tbv_y^2\rangle=\langle\tbv^2\rangle/2$.

\section{Velocity autocorrelation}

\begin{figure*}
    \centering
    \includegraphics[width=\linewidth]{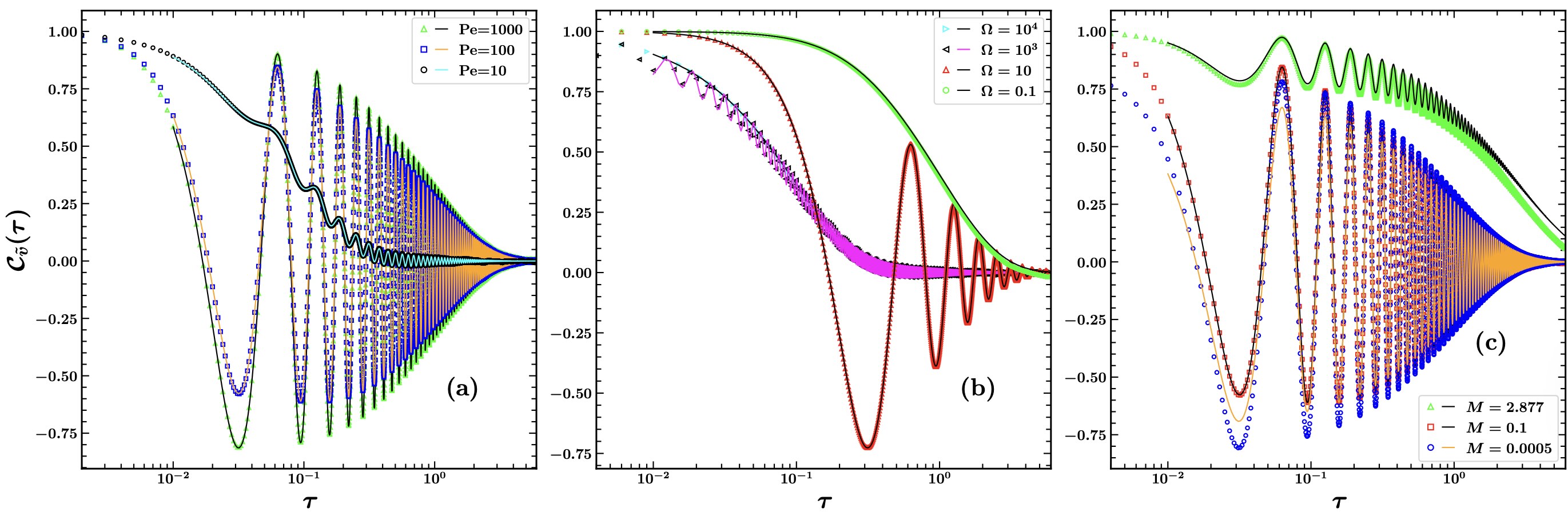}
    \caption{Normalised velocity autocorrelation
    $\mathcal{C}_{\tbv}(\tau)=\langle \tbv(\ttt+\tau)\!\cdot\!\tbv(\ttt)\rangle_{\rm st}/\langle \tbv^2\rangle_{\rm st}$
    for a cABP, shown versus lag $\tau$ for:
    (a) varying $\Pe$ at fixed $\Omega=100$, $M=0.1$;
    (b) varying $\Omega$ at fixed $\Pe=100$, $M=0.1$;
    (c) varying $M$ at fixed $\Omega=100$, $\Pe=100$.
    Curves: analytics from Eq.~\eqref{eq:velcorr_analytical}; symbols: simulation.}
    \label{velCorr}
\end{figure*}

We consider the vector velocity autocorrelation (VACF) and use steady-state normalisation:
\begin{equation}
\mathcal{C}_{\tbv}(\tau)
= \frac{\langle \tbv(\ttt+\tau)\!\cdot\!\tbv(\ttt)\rangle_{\rm st}}
        {\langle \tbv^2\rangle_{\rm st}},
\end{equation}
where $\langle \tbv^2\rangle_{\rm st}$ follows from Eq.~\eqref{eq:MSV_steady}.
Using Eq.~\eqref{first_moment_velocity}, we obtain
\begin{align}
\begin{aligned}
\langle \tbv(\ttt+\tau)\!\cdot\!\tbv(\ttt)\rangle_{\rm st}
&= \frac{e^{-\tau/M}\!\left(M^4(\Omega^2+1)\big(\Pe^2+2\Omega^2+2\big)
- M^2\big(\Pe^2-4\Omega^2+4\big)+2\right)}
        {\,M^5(\Omega^2+1)^2+2M^3(\Omega^2-1)+M\,}
\\
&\quad
+ \frac{\Pe^2 e^{-\tau}}{\,M^4(\Omega^2+1)^2+2M^2(\Omega^2-1)+1\,}
\Big(\big(M^2(\Omega^2-1)+1\big)\cos(\Omega\tau)
\\
&\qquad\qquad\qquad\qquad\qquad\qquad
+ 2M^2\Omega\,\sin(\Omega\tau)\Big).
\end{aligned}
\label{eq:velcorr_analytical}
\end{align}
As before, $\mathcal{C}_{\tbv}(\tau)$ is a superposition of two exponential decays with distinct time
scales: a purely inertial contribution 
and a chiral–rotational contribution
that is modulated at frequency $\Omega$. In Fig.~\ref{velCorr} we have plotted the velocity autocorrelation function to investigate the effects of activity ($\Pe$), chirality ($\Omega$), and inertia ($M$) on the time-dependent behavior of $\mathcal{C}_{\tbv}(\tau)$.

\paragraph{Behaviour across parameters.}
\begin{itemize}
\item[(a)] {Varying activity \(\Pe\) (fixed \(\Omega=100\), \(M=0.1\)).}
For small \(M\), the $e^{-\tau/M}$ piece decays very rapidly; the VACF is dominated by the
$e^{-\tau}$ term and shows clear damped oscillations with amplitude $\propto \Pe^2$ (Fig.~\ref{velCorr}(a)).
Larger \(\Pe\) increases the oscillation amplitude without changing the decay rate.

\item[(b)] {Varying chirality \(\Omega\) (fixed \(\Pe=100\), \(M=0.1\)).}
For very small \(\Omega\), $\sin(\Omega\tau)\!\approx\!\Omega\tau$, $\cos(\Omega\tau)\!\approx\!1$,
so oscillations are weak. For intermediate \(\Omega\) (e.g.\ 10–100) the modulated term is
pronounced (Fig.~\ref{velCorr}(b)). At very large \(\Omega\), the prefactor in the second term’s denominator grows
$\sim \Omega^2$, so the oscillation amplitude is suppressed and $\mathcal{C}_{\tbv}$ becomes
nearly purely exponential.

\item[(c)] {Varying mass \(M\) (fixed \(\Omega=100\), \(\Pe=100\)).}
As \(M\) increases, the inertial contribution $e^{-\tau/M}$ decays more slowly and contributes
a longer tail, while the effective amplitude of the oscillatory part decreases (roughly as a
low-pass filter with cutoff \(1/M\)). Hence, one observes a slower overall decay with weaker oscillations for larger \(M\) (Fig.~\ref{velCorr}(c)).
\end{itemize}

\paragraph{Limits.}
From Eq.~\eqref{eq:velcorr_analytical}:
\[
\begin{aligned}
&\text{(i) } M\!\to\!0: 
\langle \tbv(\ttt+\tau)\!\cdot\!\tbv(\ttt)\rangle_{\rm st}
\simeq
\Pe^2 e^{-\tau}\cos(\Omega\tau)+\frac{2 e^{-\frac{\tau }{M}}}{M},
\\
&\text{(ii) } \Omega\!\to\!0: 
\langle \tbv(\ttt+\tau)\!\cdot\!\tbv(\ttt)\rangle_{\rm st}
\simeq
\frac{\left(M \left(\Pe^2+2\right)-\frac{2}{M}\right) e^{-\frac{\tau }{M}}-\Pe^2 e^{-\tau }}{M^2-1},
\\
&\text{(iii) } M\!\to\!\infty: 
\langle \tbv(\ttt+\tau)\!\cdot\!\tbv(\ttt)\rangle_{\rm st}
\simeq \frac{\Pe^2+2 \Omega ^2+2}{M \left(\Omega ^2+1\right)},
\\
&\text{(iv) } \Omega\!\to\!\infty: 
\langle \tbv(\ttt+\tau)\!\cdot\!\tbv(\ttt)\rangle_{\rm st}
\simeq \frac{2 e^{-\frac{\tau }{M}}}{M}+\cos(\tau  \Omega)\frac{\Pe^2 e^{-\tau }}{M^2 \Omega ^2}+\sin(\tau  \Omega)\frac{2 \Pe^2 e^{-\tau }}{M^2 \Omega ^3}.
\end{aligned}
\]
illustrating the crossover from rotation-dominated to inertia-dominated relaxation.
(Here we displayed leading orders to show trends; full expressions follow directly from
Eq.~\eqref{eq:velcorr_analytical}.) The last term in (i) reduces to $2 \delta(\tau)$ reproducing the velocity autocorrelation function for overdamped chiral ABP.

\section{Mean squared displacement}
 To evaluate mean squared displacement, we set $\psi=\tbr^2$. The moment-generating equation(eq.~[\ref{ME_inertia}]) gives $\langle \tbr^2\rangle_s=2\langle\tbv\cdot\tbr\rangle_s/s$. $\psi=\tbv\cdot\tbr$ gives $\left(s+\frac{1}{M}\right)=\langle \tbv^2\rangle_s+\frac{2 \Pe}{M}\langle\tbr\cdot\bu\rangle_s$. In the next step, we set $\psi=\tbr\cdot\bu$ and the moment generating equation gives $(s+1)\langle \tbr\cdot\bu\rangle_s=\langle\tbv\cdot\bu\rangle_s+\Omega\langle \partial_\phi~\tbr\cdot\bu\rangle$. Following the same procedure we get $(s+1+\frac{\Omega^2}{s+1})\langle \tbr\cdot\bu\rangle_s=\langle\tbv\cdot\bu\rangle_s+\frac{\Omega}{s+1}\langle\partial_\phi\tbv\cdot\bu\rangle_s$. In the previous section we have already evaluated $\langle\tbv\cdot\bu\rangle_s$ and $\langle\partial_\phi\tbv\cdot\bu\rangle_s$. Using that and taking inverse Laplace transform of $\langle \tbr^2\rangle_s$ the exact expression of mean squared displacement is evaluated as,
       \begin{align}
       \begin{aligned}
          & \langle \tbr^2\rangle =2\ttt\frac{ \left(\Pe^2+2 \Omega ^2+2\right)}{\Omega ^2+1}+e^{-\frac{(M+1) \ttt}{M}}2 M \Bigg[\mathcal{A}_1+
          \frac{\mathcal{C}_{1n}}{\mathcal{C}_{1d}}\cos{\Omega\ttt}+\frac{\mathcal{S}_{1n}}{\mathcal{S}_{1d}}\sin{\Omega\ttt}\Bigg] +2 M e^{-\frac{\ttt}{M}}
          \\&\Big[\frac{(M-1) M \Pe (\Pe-\tbv_0\cdot\bu_0-\tbv_0\cdot\bu_0^\perp\Omega )}{M \left(M \Omega ^2+M-2\right)+1}-M \tbv_0^2+\frac{\Pe (\Pe-\tbv_0\cdot\bu_0-\tbv_0\cdot\bu_0^\perp \Omega )}{\Omega ^2+1}+4\Big]
          \\&+M  e^{-\frac{2\ttt}{M}}\left(\frac{M \Pe (-M \Pe+2 (M-1) \tbv_0\cdot\bu_0+2 M \tbv_0\cdot\bu_0^\perp \Omega +\Pe)}{M \left(M \Omega ^2+M-2\right)+1}+M \tbv_0^2-2\right)
          \\&+e^{-\ttt}\left[\frac{\mathcal{S}_{2n}}{\mathcal{S}_{2d}}\sin{\Omega  \ttt}+\frac{\mathcal{C}_{2n}}{\mathcal{C}_{2d}}\cos{\Omega\ttt}\right]
          \\&+\Bigg[ \frac{2 \Pe \left(M^2 (-\Pe)+M (M \tbv_0\cdot\bu_0+M \tbv_0\cdot\bu_0^\perp \Omega +\tbv_0\cdot\bu_0)+\Pe\right)}{(M+1) \left(\Omega ^2+1\right)}
          \\&+M \left(M \tbv_0^2-6\right)-\frac{4 \Pe^2}{\left(\Omega ^2+1\right)^2}-\frac{M^2 (M+1) \Pe^2}{M \left(M \Omega ^2+M+2\right)+1}\Bigg].
          \label{eq:MSD_full}
       \end{aligned}
   \end{align}
   The coefficients,
  \allowdisplaybreaks 
   \begin{align*}
     \mathcal{A}_1&=  2M\Pe\frac{M \tbv_0\cdot\bu_0^\perp \Omega }{(M+1) \left(M^2 \Omega ^2+1\right)}
     \\ \mathcal{C}_{1n}& = 2 M \Pe \Big(\Pe \left(M^2 \left(1-3 \Omega ^2\right)-1\right)
    \\&\quad + \left(\left(M +1\right)^2+M^2\Omega^2\right) \left(\tbv_0\cdot\bu_0 \left(M\left(\Omega ^2-1\right)+1\right)\right.
     \\&\quad \left. +(1-2 M) \tbv_0\cdot\bu_0^\perp \Omega \right)\Big)
     \\ \mathcal{C}_{1d}& = \left(\Omega ^2+1\right) \left(M^4 \left(\Omega ^2+1\right)^2+2 M^2 \left(\Omega ^2-1\right)+1\right)
     \\\mathcal{S}_{1n}&=2 M \Pe \Big(\Omega  \left(M^2 \Pe \left(\Omega ^2-3\right)+(2 M-1) \tbv_0\cdot\bu_0 \left(M \left(M \Omega ^2+M+2\right)+1\right) \right. 
     \\&\quad \left. + \Pe\right)+\tbv_0\cdot\bu_0^\perp \left(M \left(\Omega ^2-1\right)+1\right) \left(M \left(M \Omega ^2+M+2\right)+1\right)\Big)
     \\\mathcal{S}_{1d}&=\left(\Omega ^2+1\right) \left(M^4 \left(\Omega ^2+1\right)^2+2 M^2 \left(\Omega ^2-1\right)+1\right)
      \\ \mathcal{C}_{2n}& = 2 \Pe \Big(\Pe \left(M^2 \Omega ^2+1\right) \left((3 M-1) \Omega ^2-M+1\right)
     \\& \quad -M \left(\Omega ^2+1\right) \left(\tbv_0\cdot\bu_0 \left(M^2 \Omega ^2+1\right) \left(M \left(\Omega ^2-1\right)+1\right)\right. 
      \\&\quad \left. -(M-1) M \tbv_0\cdot\bu_0^\perp \Omega  \left(2 M \Omega ^2+1\right)\right)\Big)
     \\ \mathcal{C}_{2d}& =\left(\Omega ^2+1\right)^2 \left(M^2 \Omega ^2+1\right) \left(M \left(M \Omega ^2+M-2\right)+1\right)
     \\\mathcal{S}_{2n}&= -2 \Pe \Omega  \Big(\Pe \left(M^2 \Omega ^2+1\right) \left(M \left(\Omega ^2-3\right)+2\right) \\&\quad
      +M \left(\Omega ^2+1\right) \left((2 M-1) \tbv_0\cdot\bu_0 \left(M^2 \Omega ^2+1\right) \right.
     \\&\quad \left. + M \tbv_0\cdot\bu_0^\perp \Omega  \left(M \left(M \left(\Omega ^2-1\right)+3\right)-1\right)\right)\Big)
    \\ \mathcal{S}_{2d}&=\left(\Omega ^2+1\right)^2 \left(M^2 \Omega ^2+1\right) \left(M \left(M \Omega ^2+M-2\right)+1\right)
   \end{align*}
\begin{figure}
     \centering
     \includegraphics[width=\linewidth]{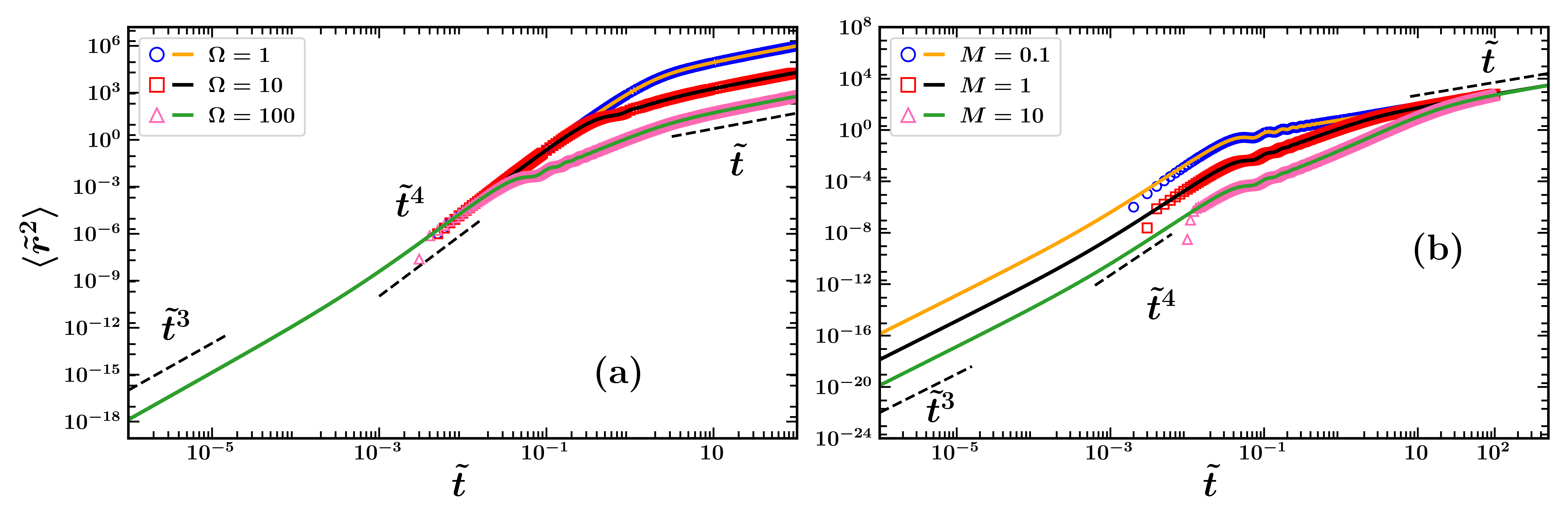}
     \caption{The mean squared displacement(MSD) is plotted as a function of time for (a) various $\Omega$ and a fixed $M=1$ and (b) various mass $M$ and a fixed $\Omega=100$. The lines represent the exact analytical expression in equation[\eqref{eq:MSD_full}] and the points are generated from the simulation. We have set initial velocity, $\tbv_0=0$. Initially, the MSD scales as $\ttt^3$ as we set $\tbv_0=0$, and then it goes as $\ttt^4$ in the later time scale. In the time limit $\ttt>>1$ it shows diffusive behavior with an effective diffusion coefficient which is a function of active velocity, $\Pe$ and chirality, $\Omega$ but independent of the inertia, $M$. In between oscillatory behavior is observed as a result of activity}
     \label{fig:msd_in}
 \end{figure}
\paragraph{Long-time diffusion.}
From Eq.~\eqref{eq:MSD_full},
\begin{equation}
\lim_{\ttt\to\infty}\frac{\langle\tbr^2(\ttt)\rangle}{4\ttt}
= \tilde D_{\rm eff}
= 1 + \frac{\Pe^2}{2(\Omega^2+1)},
\end{equation}
which is \emph{independent of} $M$ and coincides with the overdamped cABP result~\cite{pattanayak2024impact}.

\paragraph{Short-time expansion.}
For small $\ttt$,
\begin{align}
\begin{aligned}
\langle\tbr^2\rangle
&= \ttt^2\,\tbv_0^2
+ \frac{\ttt^3}{3M^2}\Big(3M\Pe\,\tbv_0\!\cdot\!\bu_0 - 3M\tbv_0^2 + 4\Big)
\\
&\quad
+ \frac{\ttt^4}{12M^3}\Big(M\Pe\big(-2(2M+5)\tbv_0\!\cdot\!\bu_0 + 3M\Omega\,\tbv_0\!\cdot\!\bu_0^\perp + 3\Pe\big)
+ 7M\tbv_0^2 - 12\Big) 
\\&\quad + \mathcal{O}(\ttt^5).
\end{aligned}
\end{align}
The ballistic $\ttt^2$ term comes from the inertial transient; active and chiral corrections enter at orders $\ttt^3$ and $\ttt^4$. Therefore, the mass affects only the short-time dynamics, while the long-time behavior is independent of it (see Fig.~\ref{fig:msd_in} for reference).

\section{Modified fluctuation--dissipation relation}

We can define a (dimensionless) kinetic temperature $\tilde T_{\rm kin}$:
\begin{equation}
k_B \tilde T_{\rm kin} = \frac{M}{2}\,\langle\tbv^2\rangle_{\rm st}
= \frac{M(M+1)\Pe^2}{2\big((M+1)^2+M^2\Omega^2\big)} + 1,
\end{equation}
using Eq.~\eqref{eq:MSV_steady}. In equilibrium (passive or $M\to 0$ overdamped limit) one has
$k_B \tilde T_{\rm kin}=1$. Chirality reduces the kinetic temperature at a fixed $(M,\Pe)$. In equilibrium, the fluctuation-dissipation (FDR) relation reads $k_B \tilde T_{\rm kin}=\tilde D_{\rm eff}$. In the presence of activity, we define the FDR violation:
\begin{equation}
\mathcal{F} \equiv \tilde D_{\rm eff} - k_B \tilde T_{\rm kin}
= \Pe^2\left(\frac{1}{2(\Omega^2+1)} - \frac{M(M+1)}{2\big(M(M\Omega^2+M+2)+1\big)}\right).
\end{equation}
Asymptotics gives:
\begin{align}
\lim_{M\to 0}\mathcal{F} &= \frac{\Pe^2}{2(\Omega^2+1)},\\
\lim_{M\to \infty}\mathcal{F} &= 0,\\
\lim_{\Omega\to 0}\mathcal{F} &= \Big(\tfrac{1}{2}-\tfrac{M(M+1)}{2\,(M(M+2)+1)}\Big)\Pe^2,\\
\lim_{\Omega\to \infty}\mathcal{F} &= 0.
\end{align}
Hence the system approaches an \emph{equilibrium-like} regime either for very heavy particles
($M\to\infty$) or at very high chirality ($\Omega\to\infty$). In the overdamped chiral ($M\to 0$) case, $\mathcal{F}\to 0$ as $\Omega\to\infty$, again recovering equilibrium-like behaviour despite persistent self-propulsion. This is different from the active Brownian particle with inertia, where the particle is, in general, driven out of equilibrium, and the equilibrium-like situation can only be achieved in the limit of very high mass.  
\begin{figure}
    \centering
    \includegraphics[width=\linewidth]{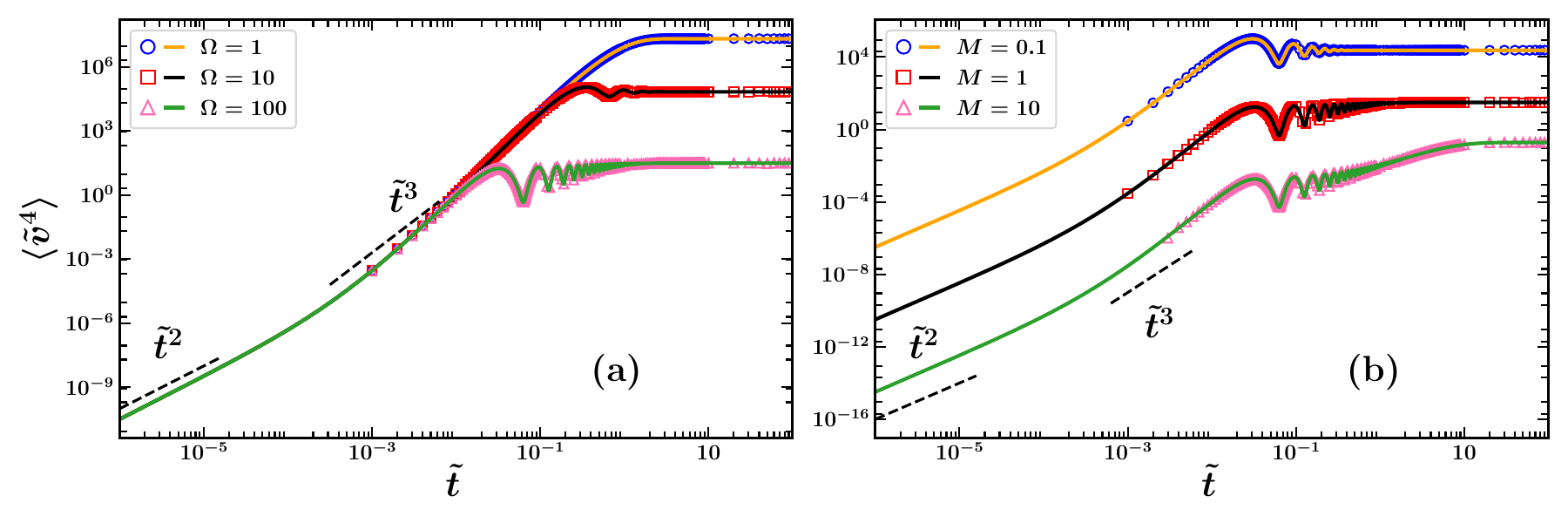}
    \caption{The temporal variation of the fourth order moment of velocity is plotted for two scenarios: (a) different values of $\Omega$ while keeping $M=1$ constant, and (b) different values of mass $M$ while keeping $\Omega=100$ constant. The lines correspond to the precise analytical formulation in Eq.~[\eqref{v4_moment}], whereas the numerical values are derived from the simulation. A starting velocity of $\tbv_0=0$ has been set up. The fourth order moment of velocity initially scales as $\ttt^2$ at $\tbv_0=0$, and then expands to $\ttt^3$ in subsequent time scales. Within the time limit $\ttt>>1$, a saturation is observed with a value of $\langle\tilde{\tbv}^4\rangle_{st}$. Periodic oscillations in the intermediate time are detected as a consequence of activity.}
    \label{v4_in}
\end{figure}

\begin{figure}
    \centering
    \includegraphics[width=0.9\linewidth]{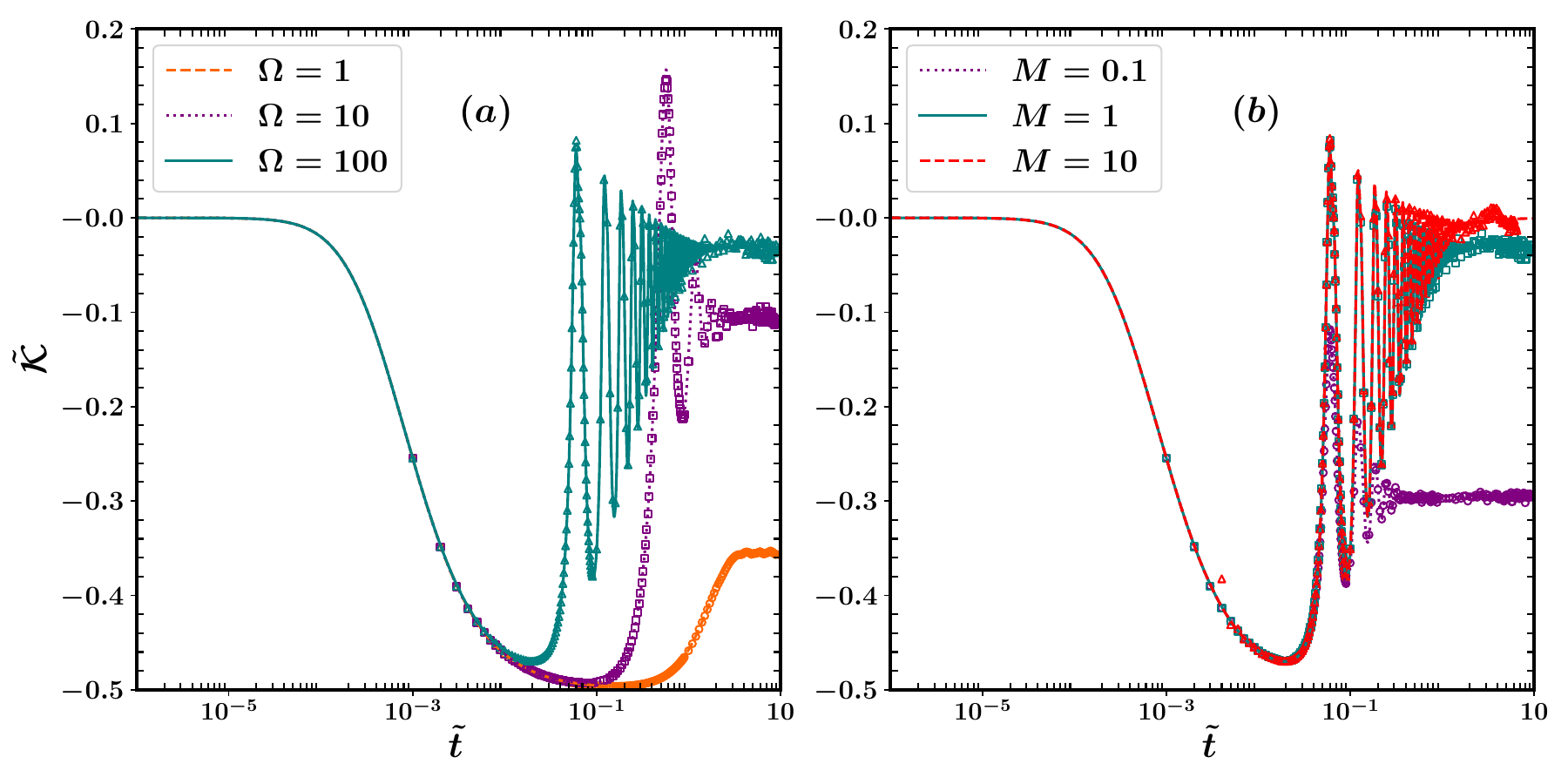}
    \caption{Excess kurtosis is plotted as a function of time for (a) varying $\Omega$ and fixed $M=1$ and $\Pe=100$, (b)  varying $M$ and fixed $\Omega=1$ and $\Pe=100$. In the short time scale, $\ttt<<1$, excess kurtosis shows Gaussian behavior and deviates from the Gaussian behavior with negative excess kurtosis at an intermediate time scale because of activity. The excess kurtosis shows oscillatory behavior in the later time crossing the zero-kurtosis line multiple times and finally in the long time limit, it saturates to its steady state value. The lines are plotted from the exact analytical expression obtained in Eq.~\eqref{kurt2d_in}. The points are generated by simulating the cABP with inertia using the Euler-Maruyama process.}
    \label{kurtosis_in}
\end{figure}

\section{Excess kurtosis}
The deviation from Gaussianity is quantified by the (dimensionless) excess kurtosis,
\begin{equation}
    \tilde{\mathcal K} \;=\; \frac{\langle \tbv^{4}\rangle}{\tilde{\mu}_{4}} \;-\; 1\,,
    \label{ex_kurt}
\end{equation}
where the fourth central moment of the 2D velocity vector is
\begin{align}
\tilde{\mu}_{4}
&= \langle \delta\tbv^{2}\rangle^{2}
  + 2\,\langle \delta \tbv_{i}\,\delta \tbv_{j} \rangle^{2}
  + 2\,\langle \delta\tbv^{2}\rangle\,\langle \tbv\rangle^{2}
  + 4\,\langle \tbv_{i}\rangle\,\langle \tbv_{j}\rangle\,\langle \delta \tbv_{i}\,\delta \tbv_{j} \rangle
  + \langle \tbv\rangle^{4},
\end{align}
with \(\delta\tbv \equiv \tbv - \langle \tbv\rangle\) and Einstein summation over Cartesian indices \(i,j\in\{x,y\}\).
Without loss of generality we set \(\tbv_{0}=\bm 0\). All second-order moments needed here were obtained in Sec.~\ref{MSV_in}.

Averaging over all initial orientations renders the steady-state velocity distribution isotropic so that
\(\langle \tbv_{i}\rangle=0\), \(\langle \tbv_{i}\tbv_{j}\rangle = \tfrac{1}{2}\langle \tbv^{2}\rangle\,\delta_{ij}\),
and \(\langle \tbv_{x}\tbv_{y}\rangle=0\). Consequently,
\begin{equation}
    \tilde{\mu}_{4} \;=\; \langle \tbv^{2}\rangle^{2} + 2\left(\tfrac{1}{2}\langle \tbv^{2}\rangle\right)^{2}
    \;=\; 2\,\langle \tbv^{2}\rangle^{2},
    \label{mu4_in_avg}
\end{equation}
and the excess kurtosis simplifies to
\begin{equation}
 \tilde{\mathcal K} \;=\; \frac{\langle \tbv^{4}\rangle}{\,2\,\langle \tbv^{2}\rangle^{2}\,} \;-\; 1.
 \label{kurt2d_in}
\end{equation}
The calculation and time-dependent expression of the fourth-order moment is given in the appendix (\ref{section_v4}). We have plotted $\tbv^4(\ttt)$ as a function of time in Fig.~\ref{v4_in}. 
\paragraph{Short-time expansion.}
Using the small-\(\ttt\) expansions of \(\langle \tbv^{2}\rangle\) and \(\langle \tbv^{4}\rangle\) (with \(\tbv_{0}=\bm 0\)),
\begin{align}
\begin{aligned}
    \lim_{\ttt\to 0} \tilde{\mathcal K}
    &= -\,\frac{\Pe^{4}}{32}\,\ttt^{2}
       + \frac{\Pe^{4}\left(3\Pe^{2}+4\right)}{192}\,\ttt^{3}\\
       &+ \frac{\Pe^{4}\,\left(120 - M^{2}\big(135 \Pe^{4}+360 \Pe^{2}-120 \Omega^{2}+136\big)\right)}{23040\,M^{2}}\ttt^{4}
       + \mathcal{O}\!\left(\ttt^{5}\right).
\end{aligned}
\end{align}

\paragraph{Steady state and asymptotics.}
At long times \((\ttt\to\infty)\), the steady state kurtosis can be obtained as given in the Appendix~\ref{section_Kss}. 
Its limiting behaviors are
\begin{align}
   \nonumber
    \lim_{M\to 0} \tilde{\mathcal K}_{\rm st}
    &= -\frac{M^{2}\Pe^{4}}{8}
       + \frac{M^{3}\Pe^{4}\left(\Pe^{2}+2\right)}{8}- \frac{M^{4}}{96}\left(\Pe^{4}\big(9(\Pe^{2}+4)\Pe^{2}-24\Omega^{2}+28\big)\right)
    \\&+ \mathcal{O}\!\left(M^{5}\right),
     \label{kst_mass_small}
     \\
    \lim_{M\to \infty} \tilde{\mathcal K}_{\rm st}
    &= \frac{\Pe^{4}\left(5\Omega^{2}-7\right)}{M\left(\Omega^{4}+5\Omega^{2}+4\right)\left(\Pe^{2}+2\Omega^{2}+2\right)^{2}}
    + \mathcal{O}\!\left(\frac{1}{M^{2}}\right),
     \label{kst_mass_high}\\
     \nonumber\lim_{\Pe\to 0} \tilde{\mathcal K}_{\rm st}&=-\frac{\text{Pe}^4 }{8}\Bigg(\frac{M^2 (M+1)^2 (M+3) (2 M+1) (7 M+3)}{\left(M^2 \Omega ^2+(M+1)^2\right)^2 \left(M^2 \Omega ^2+(M+3)^2\right) \left(M^2 \Omega ^2+(M+2)^2\right)}\\&+\frac{M^6 \Omega ^4-2 M^4 (M+1) (M (5 M-1)-5) \Omega ^2}{\left(M^2 \Omega ^2+(M+1)^2\right)^2 \left(M^2 \Omega ^2+(M+3)^2\right) \left(M^2 \Omega ^2+(M+2)^2\right)}\Bigg) + \mathcal{O}\!\left(\Pe^{5}\right)
     \label{kst_Pe_small}
     \\
    \nonumber \lim_{\Pe\to \infty} \tilde{\mathcal K}_{\rm st}& = -\frac{(M+1)^2 (M+3) (2 M+1) (7 M+3)}{2 \left((M+1)^2 \left(M \left(M \Omega ^2+M+6\right)+9\right) \left(M \left(M \left(\Omega ^2+4\right)+4\right)+1\right)\right)}\\& +\frac{M^2 \Omega ^2 \left(M \left(M \left(-10 M+\Omega ^2-8\right)+12\right)+10\right)}{2 (M+1)^2 \left(M \left(M \Omega ^2+M+6\right)+9\right) \left(M \left(M \left(\Omega ^2+4\right)+4\right)+1\right)} +\mathcal{O}\!\left(\frac{1}{\text{Pe}^2}\right)
     \label{kst_Pe_high}
   \\ \lim_{\Omega\to 0} \tilde{\mathcal K}_{\rm st}
    &= -\,\frac{M^{2}(7M+3)\Pe^{4}}
{2\left((2M^{2}+7M+3)\left(M(\Pe^{2}+2)+2\right)^{2}\right)}
    + \mathcal{O}\!\left(\Omega^{2}\right),
    \label{kst_omega_small}\\
    \lim_{\Omega\to \infty} \tilde{\mathcal K}_{\rm st}
    &= -\,\frac{\Pe^{4}}{8 M^{2}\Omega^{4}}
       + \frac{\Pe^{4}\left(M\big(M(10M+\Pe^{2}+15)+\Pe^{2}+2\big)+2\right)}{8 M^{4}\Omega^{6}}
       + \mathcal{O}\!\left(\frac{1}{\Omega^{7}}\right).
     \label{kst_omega_high}
\end{align}
From Eq.~\eqref{kst_mass_high}, the steady-state excess kurtosis is \emph{positive} in the heavy-mass limit when
\(5\Omega^{2}-7>0\), i.e.\ for \(\Omega>\sqrt{7/5}\).

\begin{figure}
    \centering
    \includegraphics[width=\linewidth]{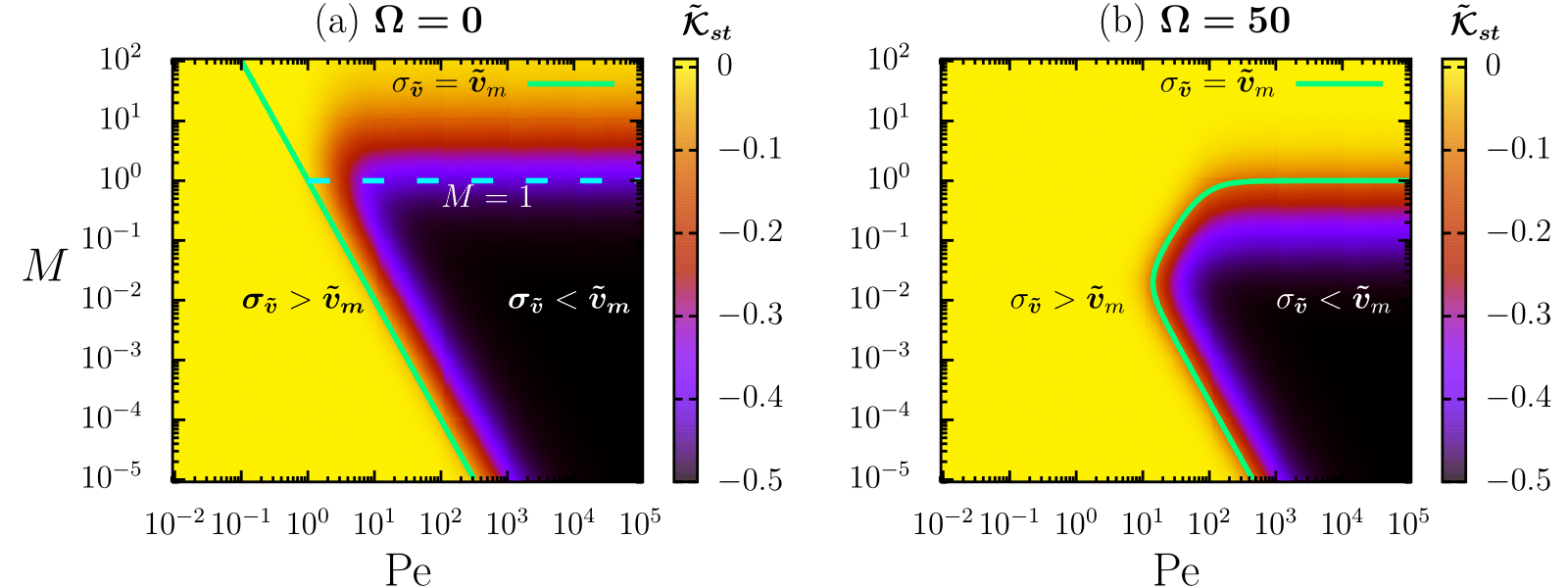}   
    \caption{
    Heatmaps of $\tilde{\mathcal K}_{\rm st}$ as functions of activity $\Pe$ (log scale) and dimensionless mass $M$ for (a) $\Omega=0$ and (b) $\Omega=50$.
    Darker shades indicate more negative $\tilde{\mathcal K}_{\rm st}$ (non-Gaussian “active” region); yellow indicates values close to $0$ (Gaussian-like region).
    Cyan solid curves show the heuristic crossover line where the thermal spread of the velocity equals the activity–induced peak speed,
    $\sigma_{\tilde v}=\tilde v_m$, with
    $\sigma_{\tilde v}\equiv \sqrt{\langle\delta \tbv_i^2\rangle}$ and
    $\tilde v_m\equiv \Pe/\sqrt{1+M^2\Omega^2}$.
    For $\Omega=0$, $\tilde v_m=\Pe$ and the (Gaussian-like)–active–(Gaussian-like) re-entrant behavior is controlled by $M$ (see text).
    The white dashed line marks $M=1$ as a visual guide.}
    \label{kurt-heatmap-massPe}
\end{figure}

\begin{figure*}
    \centering
    \includegraphics[width=\linewidth]{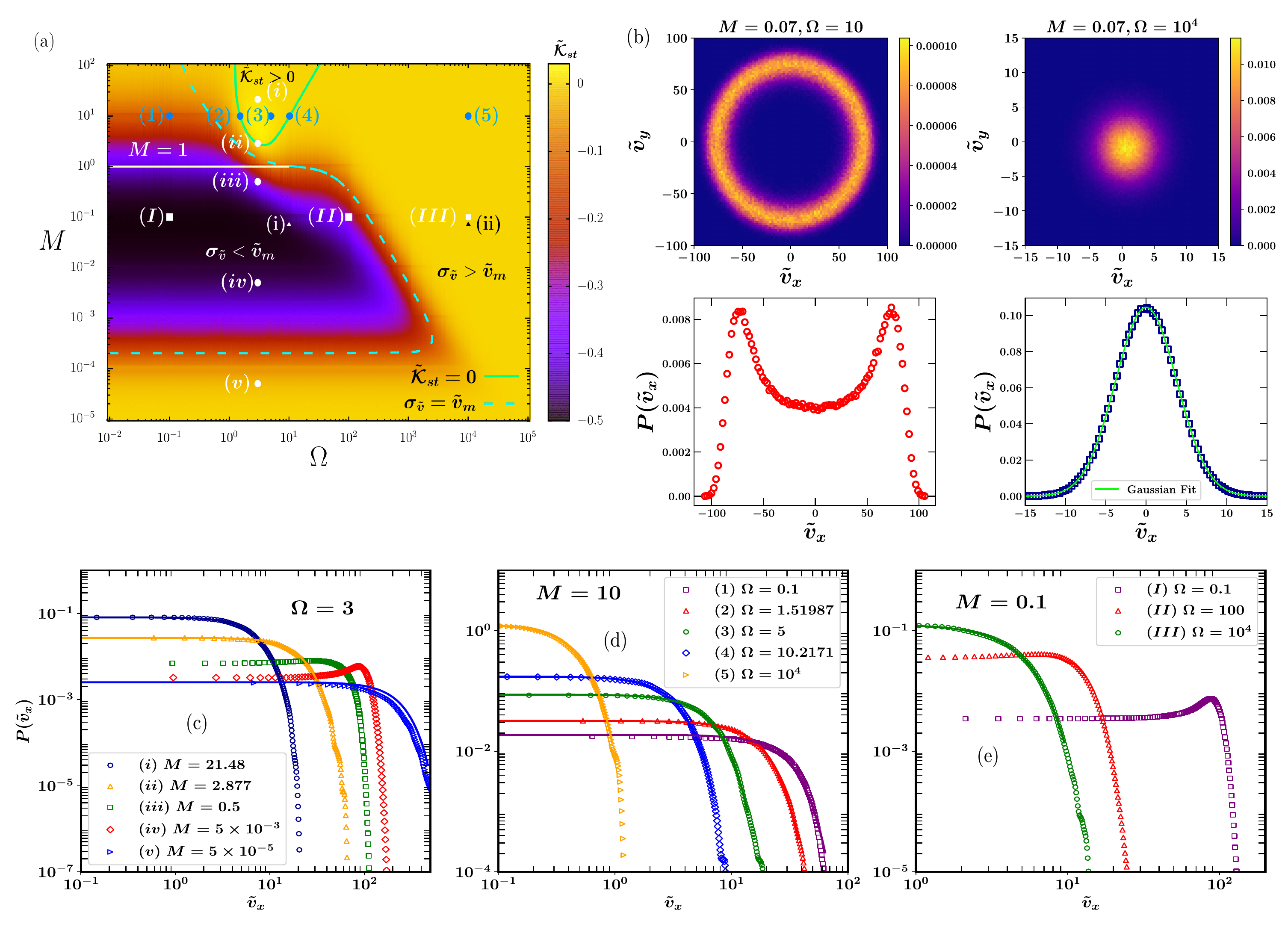}
    \caption{\small
    (a) Heatmap of $\tilde{\mathcal K}_{\rm st}$ as a function of $(M,\Omega)$ at fixed $\Pe=100$ (log axes for $M$). Darker colors denote more negative kurtosis; the thin green curve encloses a narrow region of \emph{positive} kurtosis obtained from the locus $\tilde{\mathcal K}_{\rm st}=0$.
    (b) Velocity statistics at two representative points (same $\Pe=100$): (i) “active” with non-Gaussian, off-centered distribution ($M=0.07,~\Omega=10$) and (ii) Near-Gaussian distribution ($M=0.07,~\Omega=10^4$).
    Upper row: 2D probability maps in the $(\tbv_x,\tbv_y)$ plane.
    Lower row: 1D marginals $P(\tbv_x)$ show a clear bimodal structure with peaks near $\pm\tilde v_m=\pm\,\Pe/\sqrt{1+M^2\Omega^2}$ at point (i), and a single Gaussian centered at $\tbv_x=0$ at point (ii).
    (c) $P(\tbv_x)$ for fixed $(\Pe,\Omega)=(100,3)$ and varying $M$ [points (i)–(v) from panel (a)]; (d) fixed $(\Pe,M)=(100,10)$ and varying $\Omega$ [points (1)–(5) from panel (a)]; (e) fixed $(\Pe,M)=(100,0.1)$ and varying $\Omega$ [points (I)–(III) from panel (a)].
    In dark regions of panel (a), the distributions are non-Gaussian with off-center peaks near $\tilde v_m$; in yellow regions they are well approximated by a Gaussian with zero mean and variance $\langle \tbv_x^2\rangle_{\rm st}=\langle \tbv^2\rangle_{\rm st}/2$, i.e.
    $P_{\rm G}(\tbv_x)=\exp[-\tbv_x^2/\langle\tbv^2\rangle_{\rm st}]\,/\,\sqrt{\pi\,\langle\tbv^2\rangle_{\rm st}}$ (solid lines).
    Distributions are obtained from simulations over many particles.}
    \label{kurt-heatmap-massOmega}
\end{figure*}

\subsection{Heatmap and active–(Gaussian-like) re-entrant phenomenon}
\label{section_heatmap&probability}

Figures~\ref{kurt-heatmap-massPe}(a,b) display the steady-state excess kurtosis, $\tilde{\mathcal K}_{\rm st}$, in the $(\Pe,M)$ plane for $\Omega=0$ and $\Omega=50$, respectively. For $\Omega=0$, the map shows a clear Gaussian-like ($\tilde{\mathcal K}_{\rm st}\!\approx\!0$) $\to$ active (non-Gaussian, $\tilde{\mathcal K}_{\rm st}\!<\!0$) $\to$ Gaussian-like ($\tilde{\mathcal K}_{\rm st}\to0$)) re-entrant sequence as $M$ increases at fixed $\Pe$, recovering the qualitative trends reported for achiral inertial ABPs~\cite{patel2023exact}. Physically, three regimes emerge. In the heavy-mass limit ($M\!\gg\!1$), inertial relaxation $\tau=m/\Gamma$ dominates the persistence time $\tau_r=D_r^{-1}$, so activity acts as an additional (effective) thermal forcing and the velocity statistics become Gaussian. For intermediate mass ($M\!=\!\mathcal O(1)$ and $M\leq1$), persistence dominates and produces non-Gaussian, off-centered velocity distributions whose peaks are set by the activity scale. In the light-mass limit ($M\!\ll\!1$), the spread of the velocity,
$\sigma_{\tilde v}=\sqrt{\langle\delta\tbv_i^2\rangle}\simeq \sqrt{\langle\tbv^2\rangle_{\rm st}/2}\sim M^{-1/2}$,
exceeds the activity-induced peak and the distribution again appears near-Gaussian. For sufficiently small $\Pe$ values, effects of activity and chirality become negligible on the dynamics. In this limit, the velocity distribution approaches to a Gaussian, $P(\tbv_i)= \exp\left(-\tbv_i^2/\langle\tbv^2\rangle_{st}\right)/\sqrt{\pi \langle\tbv^2\rangle_{st}}$ producing near-zero excess kurtosis.

\vspace{8pt}
\noindent
Beyond this qualitative picture, the inertial \emph{chiral} problem admits a compact, quantitative boundary condition that plays exactly the role of the length-scale comparison in trapped overdamped cABPs. The noiseless active speed (the location of the off-centered peak in the velocity PDF) is
\begin{equation}
\tilde v_m \;=\; \frac{\Pe}{\sqrt{1+M^2\Omega^2}}\,,
\end{equation}
while the steady spread of the velocity components follows from the exact MSV as
\begin{equation}
\sigma_{\tilde v} \;\equiv\; \sqrt{\frac{\langle\tbv^2\rangle_{\rm st}}{2}}
\;=\;
\sqrt{\frac{1}{2}\left[\frac{(M+1)\Pe^2}{M\,(M\Omega^2+M+2)+1}+\frac{2}{M}\right]}\,.
\end{equation}
Non-Gaussian (``active'') statistics occur when the off-centered peaks emerge from the thermal/active spread, i.e., when $v_m \;>\; \sigma_{\tilde v}$: 
\begin{equation}
\frac{\Pe^2}{1+M^2\Omega^2} \;>\; \frac{1}{2}\!\left[\frac{(M+1)\Pe^2}{M\,(M\Omega^2+M+2)+1}+\frac{2}{M}\right].
\label{eq:boundary_vm_sigma}
\end{equation}
The locus $\tilde v_m=\sigma_{\tilde v}$ (drawn as the cyan curve) tracks the bright-to-dark transition in the heatmaps and provides the inertial analogue of the radius-vs-width criteria used in confined overdamped cABPs. For $\Omega=0$, Eq.~\eqref{eq:boundary_vm_sigma} reduces to the heuristic line used previously, $\sigma_{\tilde v}\!=\!\tilde v_m\!=\!\Pe$.

\vspace{8pt}
\noindent
This criterion also explains how chirality \emph{confines} the active motion and shrinks the dark (non-Gaussian) domain in Fig.~\ref{kurt-heatmap-massPe}(b). Both $\tilde v_m$ and the large-$M$ spread,
\[
\sigma_{\tilde v}\xrightarrow[M\to\infty]{}\sqrt{\frac{1}{2M}\left(\frac{\Pe^2}{\Omega^2+1}+2\right)},
\]
decrease with $\Omega$, pushing the (Gaussian-like)–active boundary to \emph{smaller} $M$ and creating a broad high-$\Omega$ band with $\tilde{\mathcal K}_{\rm st}\!\approx\!0$. The same confinement is evident when we plot $\tilde{\mathcal K}_{\rm st}$ in the $(M,\Omega)$ plane at fixed $\Pe=100$ [Fig.~\ref{kurt-heatmap-massOmega}(a)]: a re-entrant sequence with $M$ persists, but a narrow \emph{positive}-kurtosis window appears at large $M$ and intermediate $\Omega$ (yellow strip bounded by the green curve obtained from $\tilde{\mathcal K}_{\rm st}=0$). Solving $\tilde{\mathcal K}_{\rm st}=0$ yields two positive roots which delimit that strip:
\begin{align}
\Omega_1 &= \sqrt{5 M-\frac{6}{M}+4-\frac{5}{M^2}-\mathcal{M}},\\
\Omega_2 &= \sqrt{5 M-\frac{6}{M}+4-\frac{5}{M^2}+\mathcal{M}},
\end{align}
where $\mathcal{M} = {\sqrt{M^4 (M+1)^2 \left(25 M^4-24 M^3-104 M^2-32 M+16\right)}}/{M^4}$.
 In the heavy-mass limit, $\Omega_1\!\to\!\sqrt{7/5}$ and
$\Omega_2\!=\!\sqrt{10}\,M^{1/2}+\frac{33}{10\sqrt{10}}\,M^{-1/2}+\mathcal O(M^{-3/2})$, consistent with the large-$M$ asymptote in Eq.~\eqref{kst_mass_high} predicting $\tilde{\mathcal K}_{\rm st}\!>\!0$ for $\Omega\!>\!\sqrt{7/5}$.

\vspace{8pt}
\noindent
Panels~\ref{kurt-heatmap-massOmega}(b-e) corroborate these trends using simulated velocity distributions. Dark regions (``active'') display non-Gaussian, off-centered bimodal PDFs with peaks at $\pm\tilde v_m$, whereas yellow regions are well fitted by the Gaussian reference
$P_{\rm G}(\tbv_x)=\exp[-\tbv_x^2/\langle\tbv^2\rangle_{\rm st}]\,/\,\sqrt{\pi\,\langle\tbv^2\rangle_{\rm st}}$.
The boundary~\eqref{eq:boundary_vm_sigma} provides a single, parameter-free overlay that explains the (Gaussian-like)-active-(Gaussian-like) re-entrance with $M$ and the active-to-(Gaussian-like) crossover with increasing $\Omega$, unifying the heatmap phenomenology with the underlying inertial-chiral kinematic confinement.

\section{Discussion}
Our analysis provides an exact, time-resolved characterization of a chiral active Brownian particle with finite mass, revealing how inertia and chirality conspire to shape both transient dynamics and steady-state statistics. By solving the moment-generating hierarchy, we obtained closed-form expressions for the mean velocity, velocity-orientation projections, velocity autocorrelation, MSD, MSV, and the fourth velocity moment. These results clarify several issues that have remained qualitative in prior work on inertial active matter and chiral ABPs. First, the VACF decomposes into two exponential sectors, an inertial envelope $\sim e^{-\tilde t/M}$ and a chiral envelope $\sim e^{-\tilde t}$ modulated by $\sin(\Omega\tilde t)$ and $\cos(\Omega\tilde t)$. This structure explains the observed oscillations with activity and their attenuation with mass, and it connects directly to the noise–free circular solution whose mean speed amplitude is reduced to $\Pe/\sqrt{1+M^2\Omega^2}$. Second, despite rich velocity transients, the MSD crosses over to diffusion with an effective coefficient $\tilde D_{\rm eff}=1+\Pe^2/[2(\Omega^2+1)]$ that is \emph{independent of mass}. Thus inertia reorganizes velocity statistics without renormalizing long-time transport in position space.

\vspace{8pt}
\noindent
Relative to the overdamped chiral literature and to recent theories of inertial ABPs, our modified fluctuation–dissipation relation makes the contrast sharp. Using the steady-state MSV we defined a kinetic temperature and showed that the FDR violation, $\mathcal F=\tilde D_{\rm eff}-k_B\tilde T_{\rm kin}$, vanishes in two distinct limits: $M\to\infty$ (inertial averaging) and $\Omega\to\infty$ (chiral averaging). This extends overdamped cABP results by identifying chirality as an additional route to equilibrium-like behavior, and it complements inertial, achiral analyses that recover equilibrium only at large mass. Our kurtosis maps then organize the velocity statistics into Gaussian-like and active (non-Gaussian) regions. We recover the (Gaussian-like)–active–(Gaussian-like) re-entrance with $M$ known for inertial ABPs and show that chirality \emph{shrinks} the active sector by confining the drift to $\tilde v_m=\Pe/\sqrt{1+M^2\Omega^2}$ while simultaneously reducing the heavy-mass spread. A small but robust positive-kurtosis window appears at large $M$ and intermediate $\Omega$, bounded by analytic roots of $\tilde{\mathcal K}_{\rm st}=0$ and approaching the onset $\Omega>\sqrt{7/5}$ as $M\to\infty$.

\vspace{8pt}
\noindent
These results matter beyond mathematical completeness. They suggest concrete experimental signatures in vibrobots/hexbugs, magnetically driven Janus colloids, and light-activated microflyers: VACF oscillations at frequency $\Omega$ with a mass-controlled envelope; velocity PDFs that switch from off-centered, bimodal forms (peaks at $\pm \tilde v_m$) to single-peak Gaussians as $M$ or $\Omega$ increase; and an FDR-violation curve that collapses toward zero under either heavy mass or rapid rotation. The mass-independence of $\tilde D_{\rm eff}$ offers a clean diagnostic separating velocity-sector nonequilibrium from positional diffusion.

\vspace{8pt}
\noindent
Our study has limitations. Rotational dynamics were treated as overdamped, propulsion noise was white, and hydrodynamic and interparticle couplings were neglected. These choices isolate the inertia-chirality interplay but omit effects-finite rotational inertia, colored activity, boundaries, and interactions-that can alter spectra and steady-state statistics. Future work can leverage our moment framework to include rotational inertia, active colored noise, and weak confinement, and to extend from single-particle statistics to interacting ensembles where chirality is known to suppress MIPS. Testing the predicted positive-kurtosis corridor and the dual-route FDR restoration in controlled experiments would provide stringent validation and help chart design rules for chiral microflyers and actively rotated colloids.


\appendix
\renewcommand{\thesubsection}{\thesection.\arabic{subsection}}
\renewcommand{\thesubsubsection}{\thesubsection.\arabic{subsubsection}}
\section{Noise-free trajectory} 
\begin{figure}[h]
    \centering
    \includegraphics[width=\linewidth]{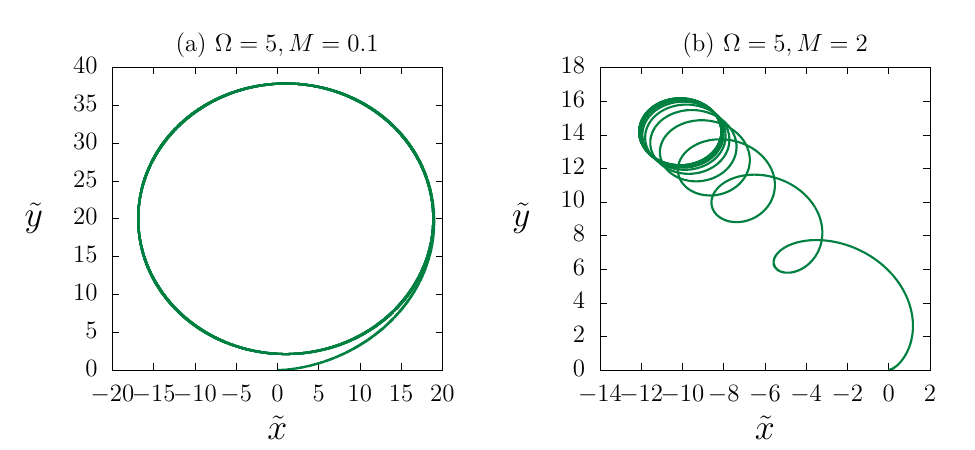}
    \caption{Noiseless trajectory is plotted for (a) $M=0.1$ and (b) $M=2$ for a constant $\Pe=100$  and $\Omega=5$. We fixed initial conditions for (a) $\phi_0=0$, $\tbv_x=10$, $\tbv_y=0$ and for (b) $\phi_0=\pi/4$, $\tbv_x=2$, $\tbv_y=0$.}
    \label{fig:noiseless_trajectory}
\end{figure}

The noiseless trajectory for two different values of $M$ are plotted in Fig.~\ref{fig:noiseless_trajectory}.
 \section{Fourth moment of velocity}\label{section_v4}
 In this section, we present all the calculations and results considering initial velocity,$\tbv_0=0$. Putting $\psi=\tbv^4$ into moments generator equation~\eqref{ME_inertia}, we get
\begin{align}
\left\langle \tbv^{4}\right\rangle_s=\frac{1}{(s+4 /M)}\left[\frac{16}{M^2} \left\langle \tbv^2\right\rangle_s+4 \frac{Pe}{M}\left\langle \tbv^2 \tbv\cdot \bu\right\rangle_{s}\right]\,.
\label{r4moment}
\end{align}
The second term can be calculated following the steps given below,

\subsection{Calculation of $\left\langle \tbv^2 \tbv\cdot \bu\right\rangle_{s} $}
\begin{align}
&\left[s+1+3/M+\frac{\Omega^{2}}{s+1+3/M}\right]\langle \tbv^2 \tbv\cdot \bu\rangle_s =\frac{8}{M^2} \left[\langle\tbv \cdot \bu\rangle_{s}+\frac{\Omega\langle \partial_\phi(\tbv\cdot \bu)\rangle_{s}}{s+1+3/M}\right]
\nonumber\\&+ \frac{\Pe}{M}\left[2 \left\langle(\tbv \cdot \bu)^{2}\right\rangle_{s} 
+\frac{\Omega\langle \partial_\phi(\tbv \cdot \bu)^{2}\rangle_{s}}{s+1+3/M}+\langle \tbv^2\rangle_{s}\right]\,.
\end{align}

\subsection{Calculation of $\left\langle(\tbv \cdot \bu)^{2}\right\rangle_{s}$}
\begin{align}
&\left[s+4 +2/M+\frac{4 \Omega^{2}}{s+4 +2/M}\right]\left\langle(\tbv\cdot \bu)^{2}\right\rangle_{s}= \frac{2 }{M^2 s}+2 \left\langle \tbv^2\right\rangle_{s} \nonumber\\
&+2 \frac{\Pe}{M}\langle\tbv\cdot \bu\rangle_{s}+\frac{2 \Omega \frac{\Pe}{M}\left\langle \partial_\phi(\tbv\cdot \bu)\right\rangle_{s}}{s+4  +2/M} +2 \Omega^{2} \frac{\left\langle \tbv^2\right\rangle_{s}}{s+4  +2/M}\,. 
\end{align}
Again,
\begin{align}
\left\langle \partial_\phi(\tbv\cdot \bu)^{2}\right\rangle_{s}  &= \frac{1}{\left(s+4 +2 /M\right)}\Big[2 \frac{\Pe}{M}\left\langle \partial_\phi(\tbv\cdot \bu)\right\rangle_{s}\nonumber\\&-4 \Omega\left\langle(\tbv\cdot \bu)^{2}\right\rangle_{z} +2 \Omega\left\langle \tbv^2\right\rangle_{s}\Big]\,.
\end{align}
Taking the inverse Laplace transform of the final expression of $\langle \tbv^4\rangle_s$ we can compute the time-dependent expression of the fourth moment of the velocity.
\begin{align}
   \langle \tbv^4(\ttt)\rangle = A_1 + \frac{A_2}{A_3} + \frac{A_4}{A_5} + \frac{A_6}{A_7} -\frac{A_8}{A_9} + \frac{A_{10}}{A_{11}}
   \label{v4_moment}
\end{align}
with
\begin{align*}
   A_1 &= -\frac{
        4 e^{-\frac{2 \ttt}{M}} \left[ M^4 \left(\Pe^2 + 2 \Omega^2 + 2\right)^2
        - M^2 \left(\Pe^4 + 4 \Pe^2 - 8 \Omega^2 + 8\right) + 4 \right]}{
        M^2 \left[ M^4 \left(\Omega^2 + 1\right)^2
        + 2 M^2 \left(\Omega^2 - 1\right) + 1 \right] }
\end{align*}
\begin{align*}
    A_2 &= e^{-\frac{4 \ttt}{M}} \Big(
    2 M^6 \left(\Omega^2 + 4\right) \left(\Pe^2 + 2 \Omega^2 + 2\right)^2  - 2 M^5 \Big[ \Pe^4 \left(2 \Omega^2 + 21\right)  
    \\& \quad + 4 \Pe^2 \left(\Omega^4 + 15 \Omega^2 + 32\right) + 48 \left(\Omega^4 + 4 \Omega^2 + 3\right) \Big]  
    \\&+ M^4 \Big[ \Pe^4 \left(\Omega^2 + 67\right)  + 8 \Pe^2 \left(20 \Omega^2 + 89\right) + 88 \Omega^4 + 768 \Omega^2 + 968 \Big] \\
    & \quad \quad  - 2 M^3 \Big[ 21 \Pe^4 + \Pe^2 \left(40 \Omega^2 + 412\right)  + 96 \left(3 \Omega^2 + 8\right) \Big] \\
    & \quad    + M^2 \Big[ 9 \Pe^4 + 408 \Pe^2  + 152 \Omega^2 + 1232 \Big] - 24 M \left(3 \Pe^2 + 20\right) + 72
    \Big) \\
   A_3 &= M^2 \Big( M^6 \left(\Omega^2 + 1\right)^2 \left(\Omega^2 + 4\right)  - 12 M^5 \left(\Omega^4 + 4 \Omega^2 + 3\right) \\& + M^4 \left(11 \Omega^4 + 96 \Omega^2 + 121\right) - 24 M^3 \left(3 \Omega^2 + 8\right)  \\&+ M^2 \left(19 \Omega^2 + 154\right) - 60 M + 9 \Big)
\end{align*}
\begin{align*}
    A_4 &= 2 M^6 \left(\Omega ^2+4\right) \left(\text{Pe}^2+2 \Omega ^2+2\right)^2+M \left(72 \text{Pe}^2+480\right)+72\\&+2 M^5 \left(\text{Pe}^4 \left(2 \Omega ^2+21\right)+4 \text{Pe}^2 \left(\Omega ^4+15 \Omega ^2+32\right)+48 \left(\Omega ^4+4 \Omega ^2+3\right)\right)\\&+M^4 \left(\text{Pe}^4 \left(\Omega ^2+67\right)+8 \text{Pe}^2 \left(20 \Omega ^2+89\right)+88 \Omega ^4+768 \Omega ^2+968\right)\\&+2 M^3 \left(21 \text{Pe}^4+8 \left(5 \text{Pe}^2+36\right) \Omega ^2+412 \text{Pe}^2+768\right)\\&+M^2 \left(9 \text{Pe}^4+408 \text{Pe}^2+152 \Omega ^2+1232\right)
\end{align*}
\begin{align*}
    A_5 & =M^2 \left(M^2 \Omega ^2+(M+1)^2\right) \left(M^2 \Omega ^2+(M+3)^2\right) \left(M^2 \Omega ^2+(2 M+1)^2\right)
\end{align*}

\begin{align*}
    A_6& = 4 \Pe^2 e^{-\frac{(M+3) \ttt}{M}} \cos(\Omega \ttt) \Bigg[ M^8 \left(2 \Pe^2 \Omega ^6-24 \Pe^2 \Omega ^4+6 \Pe^2 \Omega ^2+4 \Omega ^8+36 \Omega ^6-4 \Omega ^4-36 \Omega ^2\right)
    \\&+M^7 \left(-\Pe^2 \Omega ^6+6 \Pe^2 \Omega ^4-17 \Pe^2 \Omega ^2+192 \Omega ^4-192 \Omega ^2\right)+M^2 \left(39 \Pe^2+112 \Omega ^2+148\right)\\&+M^6 \left(19 \Pe^2 \Omega ^4-114 \Pe^2 \Omega ^2+15 \Pe^2+48 \Omega ^6+220 \Omega ^4-152 \Omega ^2-36\right)\\&+M^5 \left(-11 \Pe^2 \Omega ^4+32 \Pe^2 \Omega ^2+31 \Pe^2-192 \Omega ^4+576 \Omega ^2-192\right)-3 M \left(3 \Pe^2+64\right)+36\\&+M^4 \left(56 \Pe^2 \Omega ^2-54 \Pe^2+120 \Omega ^4+332 \Omega ^2-148\right)+M^3 \left(-19 \Pe^2 \Omega ^2-22 \Pe^2-384 \left(\Omega ^2-1\right)\right)\Bigg]
    \\&+4 \Pe^2 M^2\Omega e^{-\frac{(M+3) \ttt}{M}} \sin(\Omega \ttt) \Bigg[ 72+M^6 \left(12 \text{Pe}^2 \Omega ^4-20 \text{Pe}^2 \Omega ^2+8 \Omega ^6+80 \Omega ^4+72 \Omega ^2\right)\\&+M^5 \left(-5 \text{Pe}^2 \Omega ^4+10 \text{Pe}^2 \Omega ^2-9 \text{Pe}^2+384 \Omega ^2\right)+M^4 \left(80 \text{Pe}^2 \Omega ^2-68 \text{Pe}^2+88 \Omega ^4+448 \Omega ^2+72\right)\\&+M^3 \left(-50 \text{Pe}^2 \Omega ^2-26 \text{Pe}^2-384 \left(\Omega ^2-1\right)\right)+4 M^2 \left(29 \text{Pe}^2+38 \Omega ^2+92\right)-M \left(45 \text{Pe}^2+384\right)\Bigg]
\end{align*}

\begin{align*}
    A_7 &= M \left(M^2 \Omega ^2+1\right) \left(\left(M-1\right)^2+M^2\Omega^2\right) \left(\left(M+1\right)^2+M^2\Omega^2\right) \left(\left( M+3\right)^2+M^2\Omega^2\right) \\&\left(\left(3M-1\right)^2+M^2\Omega^2\right)
\end{align*}

\begin{align*}
    A_8 &= 4\Pe^2 e^{-\frac{(M+1) \ttt}{M}} \cos(\Omega\ttt)\Bigg[  M^8 \left(2 \left(\Pe^2+18\right) \Omega ^6-24 \Pe^2 \Omega ^4+6 \Pe^2 \Omega ^2+4 \Omega ^8-4 \Omega ^4-36 \Omega ^2\right)
    \\&+ M^7 \left(\Pe^2 \Omega ^6-6 \Pe^2 \Omega ^4+17 \Pe^2 \Omega ^2-192 \Omega ^4+192 \Omega ^2\right)+ M^2 \left(39 \Pe^2+112 \Omega ^2+148\right)\\&+ M^6 \left(19 \Pe^2 \Omega ^4-114 \Pe^2 \Omega ^2+15 \Pe^2+48 \Omega ^6+220 \Omega ^4-152 \Omega ^2-36\right)\\&+ M^5 \left(11 \Pe^2 \Omega ^4-32 \Pe^2 \Omega ^2-31 \Pe^2+192 \Omega ^4-576 \Omega ^2+192\right)+ M \left(9 \Pe^2+192\right)+36\\&+ M^4 \left(56 \Pe^2 \Omega ^2-54 \Pe^2+120 \Omega ^4+332 \Omega ^2-148\right)+ M^3 \left(19 \Pe^2 \Omega ^2+22 \Pe^2+384 \left(\Omega ^2-1\right)\right)\Bigg]
    \\&+ 4\Pe^2 e^{-\frac{(M+1) \ttt}{M}} \sin(\Omega\ttt)\Bigg[ M^8 \Omega  \left(12 \Pe^2 \Omega ^4-20 \Pe^2 \Omega ^2+8 \Omega ^6+80 \Omega ^4+72 \Omega ^2\right)\\&+ M^7 \Omega  \left(5 \Pe^2 \Omega ^4-10 \Pe^2 \Omega ^2+9 \Pe^2-384 \Omega ^2\right)+ M^6 \Omega  \left(80 \Pe^2 \Omega ^2-68 \Pe^2+88 \Omega ^4+448 \Omega ^2+72\right)\\&+ M^5 \Omega  \left(50 \Pe^2 \Omega ^2+26 \Pe^2+384 \left(\Omega ^2-1\right)\right)+4 M^4 \Omega  \left(29 \Pe^2+38 \Omega ^2+92\right)\\&+ M^3 \left(45 \Pe^2+384\right) \Omega +72 M^2 \Omega\Bigg]
\end{align*}

\begin{align*}
    A_9 &= M \left(M^2 \Omega ^2+1\right) \left( \left(M -3\right)^2+M^2\Omega^2\right) \left( \left(M -1\right)^2+M^2\Omega^2\right) \left( \left(M +1\right)^2+M^2\Omega^2\right) \\&\left( \left(3 M +1\right)^2+M^2\Omega^2\right)
\end{align*}

\begin{align*}
A_{10} &= 2 \text{Pe}^4 e^{-\left(\frac{2}{M}+4\right) \ttt}\bigg(\left(M^4 \left(\Omega ^4-37 \Omega ^2+36\right)+M^2 \left(2 \Omega ^2-13\right)+1\right) \cos (2 \Omega \ttt)\\&+10 M^2 \Omega  \left(M^2 \left(\Omega ^2-6\right)+1\right) \sin (2  \Omega \ttt)\bigg).
\end{align*}

\begin{align*}
   A_{11} &= \left(M^2 \Omega ^2+(1-2 M)^2\right) \left(M^2 \Omega ^2+(2 M+1)^2\right) \left(M^2 \Omega ^2+(1-3 M)^2\right) \left(M^2 \Omega ^2+(3 M+1)^2\right).
\end{align*}
In the long time limit, $\langle \tbv^4\rangle(\ttt)$ approaches to a steady-state constant value.
\begin{align}
\begin{aligned}
   \langle \tbv^4\rangle_{st} &=\lim_{\ttt\to\infty} \langle \tbv^4\rangle =\frac{8}{ M^2} +\frac{(M+1) \text{Pe}^2 \left((M (M+4)+2) \text{Pe}^2+16 (M+2)\right)}{2 M (M+2) \left(M \left(M \Omega ^2+M+2\right)+1\right)}\\&+
   \frac{(2 M+1) \text{Pe}^4}{(M-2) M \left(M \left(M \left(\Omega ^2+4\right)+4\right)+1\right)}-\frac{M^2 (M+3) \text{Pe}^4}{2 \left(M^2-4\right) \left(M \left(M \Omega ^2+M+6\right)+9\right)}
\end{aligned}
\end{align}
\section{Steady-state excess kurtosis of velocity distribution}\label{section_Kss}
In the steady state, $\ttt\to\infty$, the excess kurtosis takes the form,
    \begin{align}
   \tilde{\cal{K}}_{st}= \frac{M^2 \Pe^4 \left(-M^4 \Omega ^4+2 M^2 (M+1) (M (5 M-1)-5) \Omega ^2-(M+1)^2 (M+3) (2 M+1) (7 M+3)\right)}{2 \left(M \left(M \Omega ^2+M+6\right)+9\right) \left(M \left(M \left(\Omega ^2+4\right)+4\right)+1\right) \left(M \left(M \left(\Pe^2+2 \Omega ^2+2\right)+\Pe^2+4\right)+2\right)^2} 
\end{align}
\begin{figure}[h]
     \centering
     \includegraphics[width=0.49\linewidth]{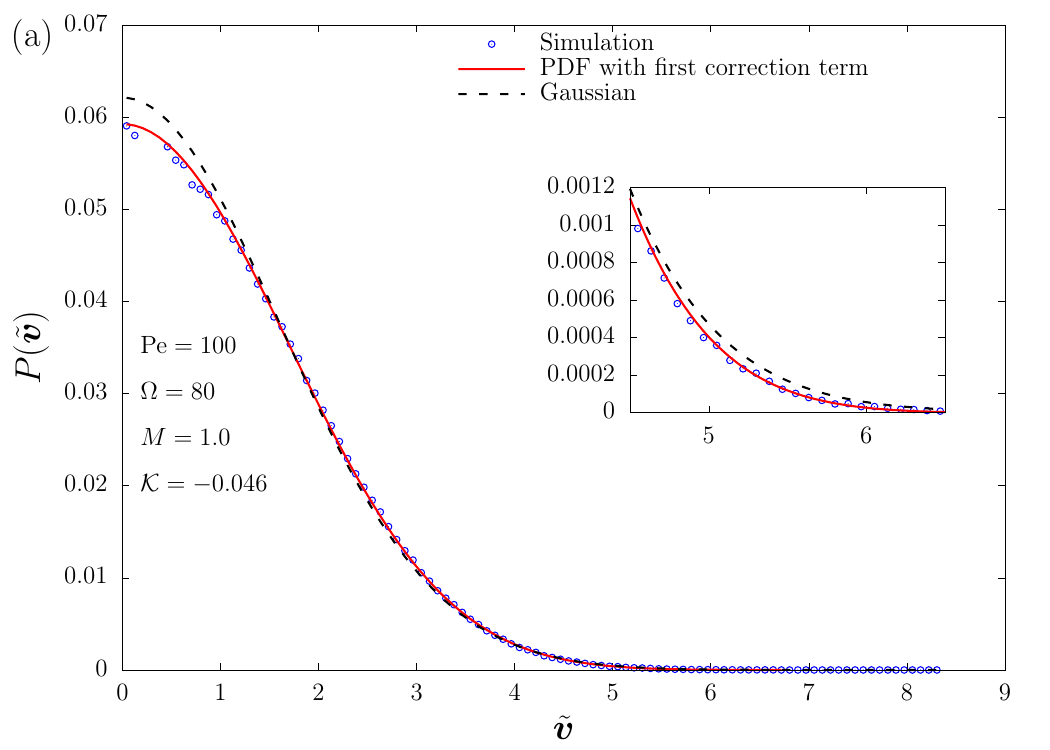}
     \includegraphics[width=0.49\linewidth]{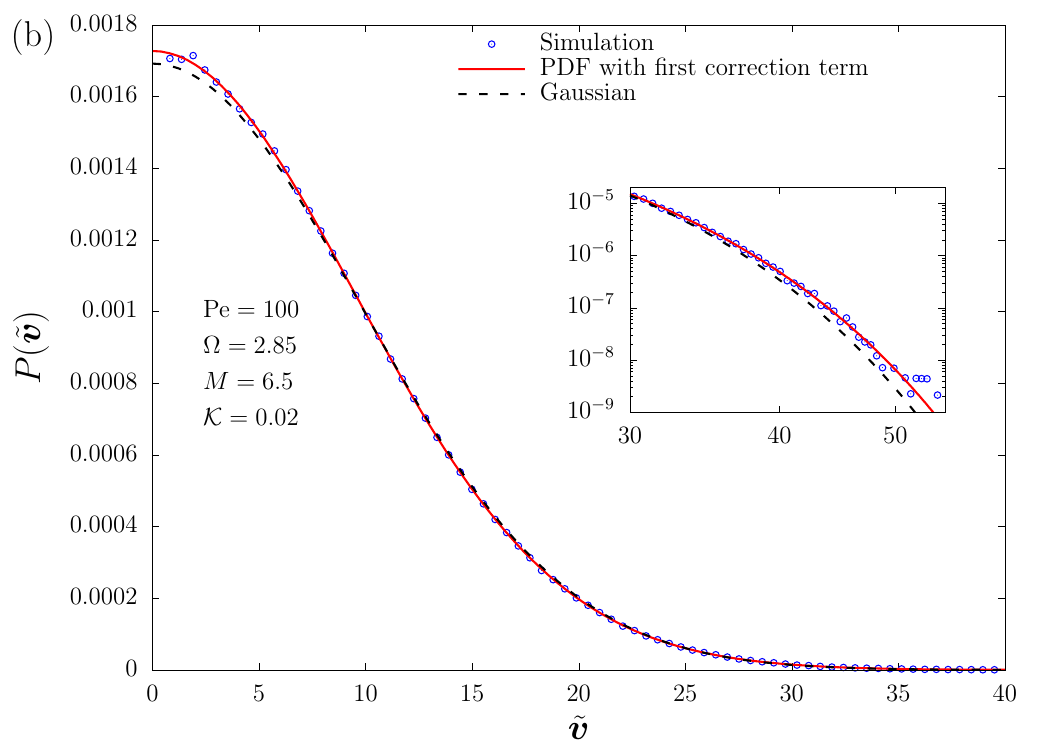}
     \caption{Velocity probability distributions obtained by simulations(denoted by points) are compared with Gaussian distributions(black, dashed lines) and distribution functions with correction term(red, solid lines), given by Eq.~\ref{corrected_prob_dist}. In the inset we show the light-tailed behavior for negative kurtosis(Fig.(a)) and heavy-tailed behavior for positive kurtosis(Fig. (b))}
     \label{fig:exact_probability}
 \end{figure}
 \section{Expression of the velocity probability distribution with first order correction -deviation from Gaussian}
 The moment generating function(MGF) is defined as the Fourier transform of the probability distribution function,
 \begin{align}
     P_k(\tbk)= \frac{1}{2\pi}\int d\tbv \exp[-i \tbk\cdot\tbv]P(\tbv)
 \end{align}
 Looking at the probability distribution heatmaps, We consider the velocity probability distribution function, in the steady state is spherically symmetric[See figure[\ref{kurt-heatmap-massOmega}]b]. Hence we can express as $P(\tbv) = P(\lvert \tbv\rvert)P(\phi)$ with $\lvert \tbv\rvert= \sqrt{\tbv_x^2+\tbv_y^2}$ and $\phi =\tan^{-1}\frac{\tbv_y}{\tbv_x}$. Considering $P(\phi)$ is uniform normalization over $\phi$ gives,
 \begin{align}
     P(\tbv)=\frac{1}{2 \pi} P(\lvert \tbv\rvert)
 \end{align}
The moment generating function takes the form, 
\begin{align}
\begin{aligned}
    P_k(\tbk) &= \frac{1}{4 \pi^2}\int_0^\infty\int_0^{2 \pi} P(\lvert \tbv\rvert)  \exp[{-i ~\lvert\tbk\rvert ~\lvert \tbv\rvert \cos(\phi)}] ~\lvert \tbv\rvert ~d\lvert \tbv\rvert ~d\phi
    \\& = \frac{1}{4 \pi^2}\int_0^\infty\int_0^{2 \pi} P(\lvert \tbv\rvert)\left(1+\sum_{n=1}^\infty\frac{(-i ~\lvert \tbk\rvert ~\lvert \tbv\rvert \cos(\phi))^n}{n!}\right)\lvert \tbv\rvert ~d\lvert \tbv\rvert ~d\phi
    \\& = \frac{1}{4 \pi^2}\int_0^\infty\int_0^{2 \pi} \lvert \tbv\rvert ~d\lvert \tbv\rvert ~d\phi P(\lvert \tbv\rvert)
    \\&\left(1-i ~\lvert \tbk\rvert ~\lvert \tbv\rvert \cos(\phi)-\frac{\lvert \tbk\rvert^2 ~\lvert \tbv\rvert^2 \cos^2(\phi)}{2}+i \frac{\lvert \tbk\rvert^3 ~\lvert \tbv\rvert^3 \cos^3(\phi)}{6}-\frac{\lvert \tbk\rvert^4 ~\lvert \tbv\rvert^4 \cos^4(\phi)}{24}+...\right)
    \\&= \frac{1}{2 \pi}\left(1 - \frac{\lvert \tbk\rvert^2}{4}\langle \tbv^2\rangle + \frac{\lvert \tbk\rvert^4}{64}\langle \tbv^4\rangle+...\right)
    \label{MGF}
\end{aligned}
\end{align}
The terms with odd $n$'s vanish. The terms $\langle \tbv^{2n}\rangle$ are even order moments of the velocity probability distribution. We can define the cumulant generating function as,
\begin{align}
    Q(\tbk) = \log_e P_k(\tbk)
\end{align}
Using Eq.[\ref{MGF}] in the expression of $Q$, we obtain 
\begin{align}
   Q(\tbk) = -\log_e(2 \pi)+\left[-\frac{\lvert \tbk\rvert^2}{4}\langle \tbv^2\rangle+\frac{\lvert \tbk\rvert^4}{32}\langle\tbv^2\rangle^2\mathcal{K}+....\right] 
\end{align}
where $\mathcal{K}=\frac{\langle \tbv^4\rangle}{2 \langle \tbv^2\rangle^2}-1$ is the excess kurtosis. Now, the moment generating function becomes
\begin{align}
   P_k(\tbk)=\frac{1}{2 \pi}\exp{\left[-\frac{\tbk^2\langle \tbv\rangle^2}{4}\right]} \left(1+\frac{\tbk^4}{32}\mathcal{K}+...\right)
\end{align}
We truncated the series at the fourth power of $\tbk$, since only velocity moments up to the fourth order are known. Taking the inverse Fourier transform, we obtain the velocity probability distribution function with the first order correction term.
\begin{align}
    P(\tbv) =\frac{\mathcal{K}~ \exp\left[{-\frac{\tbv^2}{\langle \tbv^2 \rangle }} \right]\left(\tbv^4-4 \tbv^2 \langle \tbv^2 \rangle +2 \langle \tbv^2 \rangle ^2\right)}{2 \pi  \langle \tbv^2 \rangle ^3}+\frac{e^{-\frac{\tbv^2}{\langle \tbv^2 \rangle }}}{\pi  \langle \tbv^2 \rangle }
    \label{corrected_prob_dist}
\end{align}
 For small values of excess kurtosis, this series matches excellently with the probability distribution function obtained from simulation(See Fig.~\ref{fig:exact_probability}).

\section*{References}
\bibliographystyle{iopart-num}
\bibliography{Bibliography}
\end{document}